\documentstyle[amssymb,preprint,prb,aps]{revtex}

\begin{document}
\draft
\title{An Intrinsic Tendency of Electronic Phase Separation into Two
Superconducting States in La$_{2-x}$Sr$_{x}$CuO$_{4+ \delta }$}
\author{B. Lorenz$^1$, Z. G. Li$^1$, T. Honma$^{1,2}$, and P.-H. Hor$^1$}
\address{$^1$Texas Center for Superconductivity and Department of Physics, University 
\\
of Houston,\\
Houston, TX 77204-5932, USA}
\address{$^2$Department of Physics, Asahikawa Medical College, Asahikawa\\
078-8510, Japan}
\date{\today}
\maketitle

\begin{abstract}
The effect of hydrostatic pressure up to 2 GPa on the superconductiong
transitions in La$_{2-x}$Sr$_{x}$CuO$_{4+ \delta}$ is investigated. The
ambient and high pressure properties of two series of samples with x=0 and
x=0.015 and 0$<\delta<$0.1 are characterized and compared by
ac-susceptibility measurements. At ambient pressure both sets of samples fit
into the same phase diagram as a function of the total hole concentration, $%
n_h$. For $n_h<0.085$ there is a single superconducting transition ($%
T_{c}\approx 30 K$) with an unusually large pressure coefficient, $%
dT_c^{(30)}/dp\approx 10$ $K/GPa$. At higher hole density ($n_h>0.085$) a
second superconducting transition ($T_{c}\approx 15 K$) follows the first
transition upon cooling and the pressure shift of this transition is
negative, $dT_c^{(15)}/dp\approx -4$ $K/GPa$. At the boundary as the hole
density is close to 0.085 the phase separation can be induced by pressure.
The results are explained in terms of a strong correlation of the
interstitial oxygen with the hole system in the CuO-planes. Pressure,
applied at ambient temperature, causes a redistribution of holes.The mobile
oxygen dopants follow and enhance $T_c$ as well as the tendency to phase
separation. If pressure is changed at low temperature ($<100 K$) the effects
on $T_c$ and phase separations are greatly diminished because the
interstitial oxygen becomes immobile at low T. Our results indicate that the
dopant effects are important. Dopants and holes should be treated as a
single globally correlated state. When thermodynamic euqilibrium is
approached in the oxygen-doped samples, we find that there is an intrinsic
tendency of electronic phase separation of doped holes into two distinct
superconducting states.
\end{abstract}






\section{Introduction}

Superconductivity in all high $T_{c}$ cuprates is induced by carrier doping.
Except the most recent results on field-doping experiments\cite{Schoen}, the
vast majority of data available in the literature are based on chemical
doping. Unfortunately carrier doping by chemical substitution or
intercalation is always accompanied by the introduction of dopants into the
material and the higher carrier concentration requires more dopants that
inevitably affect the electronic state of the system. Even without any
external perturbations, intrinsic dopant effects cannot be ignored.
Furthermore, since carriers can be introduced either by cation and/or by
anion doping, complications such as different lattice responses to
accommodate different kinds of dopants have to be taken into account.
Cationic doping proceeds via substitution of cations with ions of different
valency. We will use the notation ''hard doping'' since the cationic dopants
occupy lattice sites and are immobile. Anionic doping, however, frequently
results in a system of mobile dopants (e.g. interstitial or chain oxygen)
that provides an additional degree of freedom and may respond to a change of
thermodynamic parameters such as temperature or pressure. We will denote the
doping with mobile anions as ''soft doping''. $La_{2-x}Sr_{x}CuO_{4+\delta }$
is one of the high $T_{c}$-compounds that allow us to continuously tune the
relative strength of ''hard-dopant'' versus ''soft-dopant'' of randomly
distributed but immobile Sr ions (concentration $x$) and mobile interstitial
oxygen ions (concentration $\delta $), respectively. Accordingly, we can
study their influences on the normal and superconducting states as a
function of hole concentration, $n_{h}$, controlled by two parameters, $x$
and $\delta $. We use the notations ''soft-dopant'' and ''hard-dopant'' here
instead of ''annealed -disorder'' and ''quenched- disorder'' to emphasize
both the intrinsic nature (e.g. dopant increases with increasing carrier
density) and the strong coupling between carriers and dopants. The crucial
parameter is the excess-oxygen doping. The high mobility of the interstitial
oxygen provides not only the opportunity to study the ''intrinsic''
electronic evolution as a function of carrier concentration but also a
number of interesting new physical phenomena. For example, several
structural transitions as well as macroscopic phase separations have been
observed in $La_{2}CuO_{4+\delta }$ with increasing oxygen content.\cite
{Hor,Vaknin,Statt,Jorgensen,Ansaldo,Hundley,Hammel,Zolliker} For $%
0.006<\delta <0.05$ a miscibility gap was observed with two different
structures (Bmab and Fmmm symmetry) coexisting at low temperatures.
Chemically, the two structures show different oxygen contents. The Bmab
structure is not superconducting and its composition is close to
stoichiometric $La_{2}CuO_{4}$ whereas the oxygen rich Fmmm phase becomes
superconducting at about 30 K.\cite
{Hor,Vaknin,Jorgensen,Ansaldo,Hammel,Zolliker} The origin of this phase
separation was attributed to the mobile interstitial oxygen since it is not
found in the cation-doped $La_{2-x}Sr_{x}CuO_{4}$ where the dopants are
immobile and ''frozen'' in their position. For $La_{2}CuO_{4+\delta }$, one
superconducting transition was observed in the miscibility gap with a $%
T_{c}\thickapprox 30K$, almost independent of $\delta $.\cite{Hor} At higher
doping level only one crystalline structure (Fmmm) was detected\cite
{Hor,Radaelli} but also evidence for the formation of superstructures by the
interstitial oxygen was reported.\cite{Statt,Radaelli,Blakeslee} Oxygen
ordering along the c-axis (staging)\cite{Blakeslee} as well as in the
ab-plane\cite{Statt,Choi} were consequently proposed. It is interesting to
note that the superconducting transition in this regime of the phase diagram
(i.e. for $\delta >0.05$) proceeds in two well-defined steps at 30 K and 15
K with no corresponding evolution of two chemical phases suggesting the
possibility of an electronic phase separation into two distinct
superconducting phases.\cite{Hor}

Different physical and chemical methods of superoxygenation of $%
La_{2}CuO_{4+\delta }$ have been employed. Originally high-pressure oxygen
annealing was used to increase the O-content up to $\delta =0.03$ in $%
La_{2}CuO_{4+\delta }$.\cite{Jorgensen,Hundley} However, electrochemical
intercalation of oxygen has been proven to be superior to all other methods
of oxidization\cite{Radaelli,Bhavaraju,Feng} because of the low reaction
temperature (close to room temperature), the higher oxygenation level ($%
\delta $) that can be achieved, and the precise measurement and control of
the doping level. We have carefully studied various electrochemical
techniques and shown that samples so prepared are close to thermodynamic
equilibrium for $0<\delta <0.05$ with a superconducting transition at $%
T_{c}\thickapprox 30\ K$ as long as the intercalation rate was low. High
intercalation rates, in particular for $\delta >0.05$, resulted in
nonequilibrium samples with $T_{c}$ up to $45$\ $K$.\cite{Bhavaraju,Feng}
However, slow electrochemical oxidation followed by thermal annealing at $%
110\ 
{{}^\circ}%
C$ allows the samples to relax close to equilibrium and $T_{c}$ decreases to
about $30\ K$. Outside the miscibility gap for $\delta >0.05$ a second
superconducting transition with $T_{c}\thickapprox 15\ K$ develops with
annealing resulting in the two-step transition as described in the previous
paragraph.\cite{Hor,Feng}

The appearance of different phases with increasing oxygen content is
obviously related to the interstitial mobile oxygen ions and cannot be
observed in the cation-doped $La_{2-x}Sr_{x}CuO_{4}$. The strong correlation
of the negatively charged $O^{2-}$ ions with the hole carriers in the Cu-O
planes is expected to result in a cooperative behavior and the interesting
features like chemical and electronic phase separation, oxygen ordering etc.
reported in recent works. However, the driving force and the microscopic
mechanism for these phenomena are still a matter of discussion. Based on a
theoretical study of the phase diagram of a two-dimensional t-J model
including the nearest neighbor Coulomb repulsion Emery et al.\cite
{Emery,Kivelson} suggested that electronic phase separation into hole rich
and hole poor phases may be a general property of 2D hole like systems. In
this case, the mobile oxygen in $La_{2}CuO_{4+\delta }$ plays an essential
role by compensating the repulsion between holes in the hole-rich regions
and facilitating the charge separation. On the other hand, there is the
possibility that the oxygen ions cause the separation into different
(oxygen-rich and -poor) phases affecting the local hole density in the Cu-O
planes. To separate and to distinguish between these two scenarios the hole
density, $n_{h}$, and the oxygen content, $\delta $, ought to be controlled
independently. This major goal can be achieved by cation/anion co-doping in
the system $La_{2-x}Sr_{x}CuO_{4+\delta }$. Starting with a fixed Sr
content, $x$, in the parent compound and tuning the oxygen doping, $\delta $%
, by electrochemical intercalation the degree of cation- and anion-doping
can be varied at will and the hole concentration can be controlled
precisely. In order to compare samples with the same hole concentrations but
different oxygen content the doping efficiency of Sr and O has to be known.
We have calibrated the doping efficiency of pure oxygen and oxygen/strontium
co-doped samples. Whereas each Sr ion adds one hole the doping efficiency of
oxygen has been shown to change with the hole density.\cite{Li} Below $%
n_{h}=0.06$ each oxygen ion dopes exactly two holes into the planes as is
expected from the formal valency $O^{2-}$. However, for $n_{h}>0.06$ the
doping efficiency of the oxygen decreases to 1.3 holes per O-ion. Hall
measurements have shown that for $n_{h}>0.06$ the additional holes partially
occupy localized states and do not increase the Hall number.\cite{Yomo} For
the Sr co-doped compound the oxygen doping efficiency reduces from 2 to 1.3
at exactly the same critical hole density, $n_{h}=0.06$, indicating that the
O doping efficiency is controlled by the total hole density, i.e. by the
electronic state. Therefore, the total hole density in $%
La_{2-x}Sr_{x}CuO_{4+\delta }$ is determined by $n_{h}=x+2\delta $ ($%
n_{h}<0.06$) and $n_{h}=0.019+1.3(\delta +x/2)$ ($n_{h}>0.06$),
respectively. \cite{Li}

Pressure is known to increase $T_{c}$ in most of the under-doped high $T_{c}$
materials. In $YBa_{2}Cu_{3}O_{7-\delta }$ it was found that the increase of 
$T_{c}$ with pressure is enhanced by pressure-induced ordering of the oxygen
ions occupying the incompletely filled chain sites.\cite{Fietz,Sade1,Sade2}
This effect is particularly large in the far under-doped region. Another
compound where pressure (applied at ambient temperature) changes the oxygen
configuration with a crucial impact on the superconducting properties is the
overdoped $Tl_{2}Ba_{2}CuO_{6+\delta }$.\cite{Sieburger,Klehe} Whereas in
the two former compounds pressure is assumed to affect the hole density in
the active CuO-layer by transferring charges between the charge reservoir
block and the CuO-planes this mechanism does not apply to the La-214 high-$%
T_{c}$ compound. In fact, from Hall measurements at high pressures it was
concluded that the average carrier density in $La_{2}CuO_{4+\delta }$ does
not change with the application of pressure.\cite{Yomo,Yomo2} The pressure
effect on superoxygenated $La_{2}CuO_{4+\delta }$ has previously been
studied with inconsistent pressure coefficients of $T_{c}$ for various
samples prepared and doped by high-pressure oxygen annealing as well as
electrochemical intercalation. \cite{Schirber} We believe that early results
were plagued by the problem of non-equilibrium samples.

We, therefore, conducted a careful investigation of the effect of
hydrostatic pressure on the $T_{c}$ and the superconducting states of the
strontium and oxygen co-doped La-214 system, $La_{2-x}Sr_{x}CuO_{4+\delta }$%
. The goal of this work is twofold: (i) Phase separation in the correlated
hole-oxygen system has already been reported under ambient conditions.\cite
{Hor} The effect of pressure on the superconducting state and the phase
diagram as a function of hole concentration is of primary interest. To probe
the underlying driving mechanism of phase separation pressure is changed at
ambient as well as at low temperature (where oxygen motion is inhibited).
(ii) The pressure effect on samples with different oxygen content, $\delta $%
, but the same hole density, $n_{h}$, are compared to extract the most
relevant parameter determining the nature of the superconducting state.
Pressure is used as a tool to characterize the electronic state at a given $%
n_{h}$. We find that samples, both under ambient and high pressures, with
different $\delta $ and $x$ behave very similarly if compared at the same
hole concentration. This indicates that the observed phase separation,
although supported by the mobile oxygen ions, is primarily of electronic
nature.

\section{Experimental Setup}

$La_{2-x}Sr_{x}CuO_{4}$ was prepared by standard solid state reaction from
oxides and subsequently oxidized by electrochemical intercalation as
described elsewhere.\cite{Hor,Feng,Yomo} After the electrochemical
oxidization all samples were annealed at $110\ 
{{}^\circ}%
C$ in oxygen for a period of 24 to 200 h. Samples with different Sr content, 
$x=0$ and $x=0.015$, and excess oxygen $0<\delta <0.1\ $were prepared and
investigated at ambient and high pressure. DC magnetic susceptibility
measurements using a Quantum Design SQUID were employed to characterize the
superconducting transition at ambient pressure. For the high pressure
experiments the ac susceptibility technique was used. A dual coil system was
mounted to the ceramic samples and the highly sensitive low-frequency (19
Hz) LR 700 (Linear Research) mutual inductance bridge was used for the
measurements. The magnitude of the ac magnetic field was estimated to about
2 to 4 Oe. In one series of experiments hydrostatic pressure ($<2\ GPa$) was
applied at room temperature by a standard piston-cylinder beryllium-copper
clamp inserted into a liquid helium dewar for cooling. Fluorinert FC77 (3M)
was used as pressure transmitting medium. The temperature was measured using
either a k-type thermocouple inside the Teflon-pressure cell or, below 50 K,
by a germanium thermometer mounted in the beryllium-copper clamp close to
the sample position. The pressure was measured in-situ at about $7\ K$ by
monitoring the $T_{c}$-shift of a high purity (99.9999 \%) lead manometer. A
second series of experiments was conducted measuring the same samples in a
He-gas pressure cell ($p<1.2\ GPa$, Unipress) inserted into a variable
temperature cryostat (CRYO-Industries of America). Gas pressure was
transmitted from a $1.5\ GPa$ gas compressor (U11, Unipress) through a
beryllium-copper capillary (ID 0.3 mm). Pressure can be changed at any
temperature above the freezing temperature of the helium pressure medium. In
particular, pressure was changed at low temperature where the interstitial
oxygen is immobile. The temperature was controlled by pre-heated He gas in
the cryostat and measured by a thermocouple inside the pressure cell as well
as by two thermometers attached to the top and the bottom of the cell.
Special care has been taken to avoid freezing of the helium in the capillary
before it solidifies in the pressure cell. The pressure was measured using
two manometers. A manganin gauge indicated the pressure at room temperature
in the high-pressure chamber of the gas compressor. A second (semiconductor)
manometer was mounted close to the sample inside the pressure cell. As long
as the helium in the system is liquid both manometers yield identical
readings as expected. However, if the helium solidifies in the pressure cell
(this happens e.g. at $38.7\ K$ for $p=0.5\ GPa$ or at $61\ K$ for $p=1.01\
GPa$) \cite{Spain} the in-situ semiconductor gauge is extremely valuable to
detect the pressure drop during freezing and any possible blockage of the
capillary. We have found that even in the solid state of the pressure medium
the internal gauge still yields reliable pressure readings following the
known isochore of the helium solid.\cite{Spain} Data were taken during
cooling and heating through the superconducting transitions. The cooling
speed between room temperature and $150\ K$ was very slow (typically $0.4\
K/min$) in order to maintain exactly the same conditions for all
experiments. This is particularly important for doping levels where chemical
phase separations occur because it was shown that in this range the
superconducting transition temperature may depend on the cooling rate.\cite
{Hor2} However, any kinetic process leading to phase separations slows down
at low temperature and prevents the sample to reach true thermodynamic
equilibrium at temperatures close to $T_{c}$. We have tested various cooling
procedures with speeds below $0.4\ K/min$ and found no appreciable change in
the superconducting properties. Therefore, we conclude that with the chosen
procedure the samples are as close to thermodynamic equilibrium as one can
get.

\section{Results and Discussion}

\subsection{Ambient pressure results}

The dc susceptibility, $\chi _{dc}$, was measured and compared with the real
part of the ambient pressure ac susceptibility, $\chi _{ac}$. Although we
did not attempt to extract the superconducting volume fraction from the ac
measurements we could show that the relative temperature dependence of $\chi
_{ac}$ is comparable to that of $\chi _{dc}$. The results obtained by the ac
method for $La_{2}CuO_{4+\delta }$ reproduced our earlier dc data.\cite{Hor}
A single superconducting transition ($T_{c}\thickapprox 30\ K$) observed for
low doping ($\delta <0.05$) turns into a two-step transition ($T_{c}$'s at
about $15$ and $30\ K$) upon increasing $\delta $. This effect was
interpreted as a doping-induced electronic phase separation of two intrinsic
superconducting states. At $\delta =0.082$ (corresponding to the hole
concentration $n_{h}=0.125=1/8$) the superconducting volume fraction of the $%
30\ K$ phase is greatly diminished but that of the $15\ K$ phase is not
changed.\cite{Hor3}\ This results in a fast drop of the onset-$T_{c}$ for
superconductivity from about 30 K to 15 K at this special hole concentration
in analogy to the so called 1/8 anomaly in the $La_{2-x}(Sr,Ba)_{x}CuO_{4}$%
-system. The phase diagram and some typical ac-susceptibility curves are
schematically shown in Fig. 1 as a function of $n_{h}$. Also shown in the
bottom half of the figure are the same data for the Sr co-doped system, $%
La_{1.985}Sr_{0.015}CuO_{4+\delta }$. The qualitative features of both
systems are very similar. In particular, the transition from the one-$T_{c}$
to the two-$T_{c}$ regime in the diagram and the 1/8 anomaly happen at the
same hole concentration, $n_{h}$, but at different $\delta $. Note that $%
n_{h}$ here is the ''nominal'' ''total'' carrier density determined by the
equation in the introduction. Theses results reconfirm that our samples are
indeed close to equilibrium and the single- and double-$T_{c}$ behaviors are
''intrinsic'' properties of doped holes. In the very low doping region ( $%
n_{h}$ $<0.06$) there is a major difference between the two sets of samples.
There is a single $T_{c}=15\ K$ in $La_{1.985}Sr_{0.015}CuO_{4+\delta }$
whereas the $T_{c}$ of $La_{2}CuO_{4+\delta }$ remains at $30\ K$ even for
the smallest $\delta $. This change of $T_{c}$ is explained by the
attraction of charge carriers to the ''frozen'' Sr-ions. Because of the
random distribution of the cation-dopants the formation of $T_{c}=30\ K$
phase is inhibited. The Coulomb attraction between holes and Sr-ions results
in a more homogeneous distribution of carriers with a lower local hole
density and the intrinsic $T_{c}=15\ K$ phase is stabilized instead. In the
following sections we discuss the pressure effects on the superconducting
transitions in $La_{2-x}Sr_{x}CuO_{4+\delta }$ for various doping levels for
the two characteristic regimes in the phase diagram (Fig. 1) and for samples
with carrier density $n_{h}\thickapprox 0.085$ located right at the
transition boundary (shaded area in Fig. 1).

\subsection{Effect of High Pressure on the Superconducting Transition:
Pressure Applied at Room Temperature}

The ac-susceptibility data for two typical samples with $n_{h}\approx 0.068$
in the single $T_{c}=30\ K$ regime are displayed for different pressures in
Fig. 2 A ($La_{2}CuO_{4.04}$) and Fig. 2 B ($La_{1.985}Sr_{0.015}CuO_{4.027}$%
). The diamagnetic signal of $\chi _{ac}$ shifts to higher temperature and
allows for an accurate determination of the pressure coefficient of $T_{c}$,
defined as the onset of the diamagnetic drop. The pressure effect on $\chi
_{ac}(T)$ is reversible upon increasing and decreasing pressure and
reproducible in several pressure cycles. The pressure dependences of $T_{c}$
for both samples (open circles) and also for $La_{1.985}Sr_{0.015}CuO_{4+%
\delta }$ at higher oxygen doping, $\delta =0.040$ (triangles) and $\delta
=0.061$ (squares) are shown in Fig. 3. The increase of $T_{c}(p)$ is
non-linear for small pressure but linear at larger pressure. The pressure
coefficient (in the linear high-pressure range) is unusually large, $9$ to $%
10\ K/GPa$. This is about three to four times larger than the $dT_{c}/dp$ of
the cation-doped $La_{2-x}(Sr,Ba)_{x}O_{4}$\cite{XXX} and appears to be the
largest pressure coefficient observed in a La-214 high-$T_{c}$
superconductor. An anomalous large $dT_{c}/dp$\ (up to 6 K/GPa) was
previously observed only near the 1/8 anomaly in the La 214 system where the
ambient pressure $T_{c}$\ is very low.\cite{Liverman}\ This huge pressure
coefficient is obviously a consequence of the high mobility of the
interstitial oxygen dopants. In the under-doped region, pressure is known to
increase the transition temperature of high-$T_{c}$ superconductors, even in
systems without mobile oxygen such as $La_{2-x}(Sr,Ba)_{x}O_{4}$. For
multi-layer compounds this effect was frequently attributed to a pressure
induced increase of the average hole-density in the Cu-O planes due to a
charge transfer from a charge reservoir block to the active layer. However,
in the structure of La-214 there is no charge reservoir and the average hole
concentration does not change with pressure. One possibility to stabilize
superconductivity and to increase $T_{c}$ is the phase separation into
hole-rich and hole-poor phases. At ambient conditions, this phase separation
in $La_{2}CuO_{4+\delta }$ results in phases with low and high oxygen
content and different lattice structures for low doping ($\delta <0.05$) and
in the coexistence of two superconducting phases ($T_{c}=15\ K$ and $30\ K$)
for higher doping levels ($\delta >0.05$).\cite{Hor} Under pressure the
higher $T_{c}=30\ K$ state is favored and, with the support of the mobile
oxygen ions, $T_{c}$ may increase up to a rate of$\ 10\ K/GPa$. In contrast,
in the cation-doped system, $La_{2-x}(Sr,Ba)_{x}O_{4}$, the random
distribution of the immobile dopants prevents the hole system from phase
separation due to pinning of holes close to the randomly distributed dopant
sites. This explains the far lower pressure effect on $T_{c}$ in that
system. For the oxygen doped as well as the oxygen-Sr co-doped La-214
cuprate the pressure induces a redistribution of the strongly correlated
interstitial oxygen and hole system. This results in a stabilization of the
superconducting phase and the large $dT_{c}/dp$ observed in our experiments.
Similar effects have been reported in high-pressure experiments on
under-doped YBCO\cite{Fietz} and may be due to the same underlying
microscopic mechanisms.

In the high hole density ($n_{h}>0.084$) regime where we found two
superconducting transitions with $T_{c}=15\ K$ and $T_{c}=30\ K$ coexisting
at ambient pressure, the effect of pressure on the two transitions is of
particular interest. Fig. 4 shows the temperature dependence of the
ac-susceptibility at different pressures for two typical samples with $%
n_{h}\approx 0.115$, $La_{2}CuO_{4.108}$ and $La_{1.985}Sr_{0.015}CuO_{4.061}
$. The typical two-step characteristics of $\chi _{ac}(T)$ is preserved
under pressure but the critical temperatures for both transitions are pushed
into opposite direction, i.e. $T_{c}^{(30)}$ increases and $T_{c}^{(15)}$
decreases with pressure. Note that $T_{c}^{(30)}$ is defined as the onset of
the diamagnetic drop of $\chi _{ac}(T)$ and $T_{c}^{(15)}$ is determined
from the bending of the susceptibility curve as shown in the lower right
graph of Fig. 1 (dotted lines). Thereby, under high pressure $T_{c}^{(30)}$
has the same pressure coefficient as that of $\delta <0.05$ samples (see
e.g. Fig. 3 B) suggesting that it is the same superconducting state. $%
dT_{c}^{(15)}/dp$ is negative and of the order of $-4\ K/GPa$. We conclude
that pressure enhances the electronic phase separation between the two
superconducting phases because of a redistribution of charges and dopants in
the coupled hole-oxygen system. The $30\ K$ phase becomes more stabilized at
the expense of the $15\ K$ state. The pressure induced changes to the hole
distribution may result in a change of the superconducting volume fractions
of both phases but also in a change of the hole density of each
superconducting phase. The superconducting volume fraction cannot be
extracted from ac susceptibility measurements but it would be accessible by
measuring the dc susceptibility (Meissner effect) under high pressure.
However, a sole change of the volume fractions of both ($15\ K$ and $30\ K$)
phases should not affect the transition temperatures. Therefore, it appears
more likely that pressure changes the local hole density of the two phases,
for example, by increasing the hole concentration of the $30\ K$ phase and
reducing the hole density of the $15\ K$ phase. This might explain the
change of both $T_{c}$'s, however, the current experiments cannot provide
information about the local hole concentrations. We also cannot exclude the
possibility that the 15 K phase might be overdoped resulting in a negative
pressure coefficient of $T_{c}$ as known for many high-$T_{c}$ compounds.
However, the cation doped La-214 system was shown to exhibit a positive $%
dT_{c}/dp$ even for doping levels well above the optimum.\cite{XXX, Liverman}
As we will see in the following section the rearrangement of charges is
suppressed by freezing the interstitial oxygen ions in their ambient
pressure positions.

The pressure effect in the transition region (shaded area in Fig. 1) located
at $n_{h}\thickapprox 0.085$ is of special interest. The ambient pressure
susceptibility at this special hole density shows one single transition at $%
T_{c}\thickapprox 30K$ (dashed lines in Fig. 5). However, Fig. 5 shows that
under hydrostatic pressure this transition splits into two indicating that
the electronic phase separation can be induced by pressure. The relevant
parameter for this effect is the hole density, $n_{h}$. Figures 5 A and B
compare two data sets for $La_{2}CuO_{4+\delta }$ and $%
La_{1.985}Sr_{0.015}CuO_{4+\delta }$ at almost the same $n_{h}\thickapprox
0.083$, but different oxygen content. The pressure induced phase separation
into the $15\ K$ and $30\ K$ superconducting phases is obvious in both, the
oxygen doped and the Sr-oxygen co-doped samples. However, if both compounds
are compared at the same $\delta $ but slightly different hole density (Fig.
5 A and C), the pressure effect on the superconducting transition is
qualitatively different. The splitting into two transitions is induced by
pressure in $La_{1.985}Sr_{0.015}CuO_{4.04}$ but not in $La_{2}CuO_{4.04}$.
This result indicates that the separation into two $T_{c}$'s is hole driven.
If those two $T_{c}$'s were due to two distinct chemical phases then we
expect to observe the two $T_{c}$'s behavior in samples with same oxygen
content, e.g. in $La_{2}CuO_{4.04}$ too. The small increase of hole density
due to the co-doping with Sr triggers the interesting pressure effect at
this special carrier density. This observation, coupled to the fact that all
samples are close to thermal equilibrium and behave similarly at the same
carrier density, leads us to the conclusion that the hole density dictates
the superconducting properties of $La_{2-x}Sr_{x}CuO_{4+\delta }$ and indeed
the two characteristic regimes are separated by a critical hole density $%
n_{h}\thickapprox 0.085$. The observed phase separations at ambient and high
pressures are driven by the system of holes and are facilitated by the
rearrangement of the mobile interstitial oxygen ions. It should be noted
that all these effects cannot be observed in cation-doped $%
La_{2-x}(Sr,Ba)_{x}CuO_{4}$ because of the random distribution of the
immobile dopant ions. Due to the Coulomb interaction between holes and
dopant ions the hole distribution in the Cu-O planes is affected by the
dopants and electronic phase separation into hole-rich and hole-poor phases
is inhibited.

The pressure effects on the $15\ K$ transition in $%
La_{1.985}Sr_{0.015}CuO_{4+\delta }$ (for $\delta <0.024$ or $n_{h}<0.06$)
collaborates with our picture. From the former discussion it could be
expected that pressure induces the phase separation into the hole rich ($30\
K$) and hole poor (non-superconducting) phases. In fact, the
ac-susceptibility data of $La_{1.985}Sr_{0.015}CuO_{4.02}$ shown in Fig. 6
for pressures up to $1.74\ GPa$ indicate that pressure rapidly increases the 
$T_{c}$ from $15$ to above $30\ K$. The details of this transition are
clearly seen in the derivative, $d\chi _{ac}/dT$, displayed in Fig. 6B. At
ambient pressure (dashed curve in Fig. 6) $d\chi _{ac}/dT$ exhibits a single
maximum close to $15\ K$ as indication of the diamagnetic drop of $\chi
_{ac} $. This maximum is shifted to higher temperature with increasing
pressure (top part of Fig. 6B). Above $0.6\ GPa$ a shoulder appears in the
temperature dependence of $d\chi _{ac}/dT$ at about $30\ K$. This shoulder
corresponds to a small diamagnetic signal in $\chi _{ac}$ (Fig. 6A) and is
interpreted as the pressure induced growth of the hole-rich $30\ K$ phase
observed at the same hole density in $La_{2}CuO_{4+\delta }$. The sudden
appearance of this $T_{c}=30\ K$ state instead of a continuous increase of \ 
$T_{c}$ from $15\ K$ to $30\ K$ clearly demonstrate that there were two
distinct intrinsic $T_{c}$'s in this system. With further increasing
pressure this phase is growing (the shoulder extends to a major peak in $%
d\chi _{ac}/dT$, bottom part of Fig. 6B) and becomes the major
superconducting phase in the sample. There is a narrow pressure range ($p\
\thickapprox 0.9\ GPa$) where both phases coexist and the $d\chi _{ac}/dT$
curve exhibits two separated maxima. Since we have seen that the $30\ K$
transition is favored under pressure, mobile oxygen dopants now facilitate
the phase separation into hole-rich and hole-poor phases similar to the $%
La_{2}CuO_{4+\delta }$ system. This interesting pressure effect has been
observed in $La_{1.985}Sr_{0.015}CuO_{4+\delta }$ for several samples with $%
\delta <0.024$.

To prove the conclusion of this section and to collect additional evidence
for the role of interstitial oxygen we investigate the pressure effects
while oxygen ions are immobile.

\subsection{High-Pressure Effect on the Superconducting Transition: Pressure
Changed at Low Temperature}

The mobility of the interstitial oxygen is greatly reduced and the ions are
literally frozen in their positions if the temperature is reduced to below $%
100\ K$. Changing pressure at low temperature will allow us to separate the
pressure effect on the electronic state from the effect of electronic phase
separation facilitated by mobile oxygen ions. The low-temperature pressure
change can be achieved by employing a helium gas-pressure cell connected to
a gas compressor. For several typical $La_{2}CuO_{4+\delta }$ and $%
La_{1.985}Sr_{0.015}CuO_{4+\delta }$ samples the pressure effects (at $%
T<100\ K$) were investigated using three basic procedures:

(i) Pressure up to $1\ GPa$ was applied at room temperature and released in
several steps at low temperature. AC-susceptibility measurements through the
superconducting transition were conducted at each pressure step below $100\
K $.

(ii) Following the release of pressure below $100\ K$ according to procedure
(i) the samples were re-heated to room temperature so that the mobile oxygen
could re-arrange. Samples were cooled again (without changing pressure)
through the superconducting transition. The comparison of the two
measurements shows the sole effect of interstitial oxygen redistribution on $%
T_{c}$.

(iii) Samples were cooled to low temperature at ambient pressure. Pressure
was increased below 100 K and the ac-susceptibility was measured for each
pressure.

The pressure shift of $T_{c}^{(30)}$ is shown in Fig. 7 for a typical sample
in the low hole density region, $La_{1.985}Sr_{0.015}CuO_{4.027}$ ($%
n_{h}=0.064$). The pressure in this experiment was raised to $1\ GPa$ at
room temperature and released at low $T$. During cooling the pressure drops
to about $0.92\ GPa$ at the freezing temperature of the helium pressure
medium, $T_{f}=55\ K$. A further pressure decrease to $0.82\ GPa$ at$\ 35\ K$
is due to the volume reduction and the isochoric change of state of the
solid helium. The $T_{c}(0.82\ GPa)$ coincides with the previous
measurements (Fig. 2 B) and decreases with decreasing pressure at a rate of $%
2.8\ K/GPa$. This rate is less than $1/3$ of that estimated in the previous
section for the same sample (Fig. 3 B). After heating the sample to room
temperature and cooling again $T_{c}$ is further reduced by $2.2\ K$ (Inset
in Fig. 7). This additional change of $T_{c}$ is a pure effect of oxygen
redistribution since pressure was not changed. This shows that the
hole-oxygen system was frozen in a meta-stable state after the release of
pressure at low temperature and could only relax towards thermal equilibrium
after the temperature was raised high enough to allow the oxygen diffusion.
Repeating the measurement according to procedure (iii) the increase of
pressure from $p=0$ at low temperature results in a similar shift of $T_{c}$
at the rate of about $3\ K/GPa$ in good agreement with the value of $2.8\
K/GPa$ mentioned above. We conducted the same series of gas-pressure
experiments with the oxygen-doped $La_{2}CuO_{4.04}$ ($n_{h}=0.071$) and
found very similar results. Again, the observed pressure coefficient of $%
dT_{c}/dp\thickapprox $ $2.7\ K/GPa$ was strongly reduced with respect to
the value of $9.9\ K/GPa$ found in section 3.2 for this sample (see also
Fig. 3 A). It is interesting to note that the reduced $dT_{c}/dp$ is
comparable to the pressure coefficient of the cation-doped $%
La_{2-x}Ba_{x}CuO_{4}$ ($2$ to $3\ K/GPa$) and slightly larger than $%
dT_{c}/dp$ for $La_{2-x}Sr_{x}CuO_{4}$ ($\approx 1.5\ K/GPa$) for the doping
region not too close to the 1/8 anomaly.\cite{XXX, Liverman} Therefore, we
conclude that this pressure coefficient of $dT_{c}/dp\thickapprox $ $2.7\
K/GPa$ should be considered as the pressure effect of the $T_{c}^{(30)}$%
transition under the ''hard-doping'', either Sr or frozen oxygen, condition.

For the higher doping region ($n_{h}>0.085$) we choose a representative
sample, $La_{1.985}Sr_{0.015}CuO_{4.061}$. Our clamp-cell data (Section 3.2,
Fig. 4 B) indicate that pressure enhances the phase separation. This effect
cannot be observed if pressure is changed at low temperature. Fig. 8 shows
the temperature dependence of $\chi _{ac}$ at different pressures applied
according to the following procedure: Pressure was increased at room
temperature to $0.8\ GPa$. After cooling to low temperature pressure was
varied below $100\ K$ between $0.93\ GPa$ and $0.01\ GPa$. The two
superconducting transitions are clearly distinguished, however, pressure
mainly affects $T_{c}^{(30)}$ whereas $T_{c}^{(15)}$ remains almost
unchanged. The pressure coefficient $dT_{c}^{(30)}/dp=2.7\ K/GPa$ is of the
same value as that for the single transition compounds (for $n_{h}<0.085$).
After heating at $0.01\ GPa$ to room temperature the superconducting
properties of the sample changed and, in particular, $T_{c}^{(15)}$
increased by about $3.5\ K$ restoring the original ambient pressure curve $%
\chi _{ac}(T)$. Very similar data have been obtained for oxygen-doped
samples with $n_{h}>0.085$, e.g. $La_{2}CuO_{4.08}$. The freezing of the
oxygen ions at low temperature prevents the pressure induced enhancement of
electronic phase separation as observed in the experiments described in the
previous section. Thermodynamically, the system behaves more like $%
La_{2-x}(Sr,Ba)_{x}CuO_{4}$ under pressure. The major difference is the
inhomogeneous hole-oxygen distribution in $La_{2-x}Sr_{x}CuO_{4+\delta }$ we
start with at ambient conditions. The degree of phase separation already
present at ambient T is frozen and preserved at low temperature and cannot
be changed by pressure.

For $n_{h}\thickapprox 0.085$ (shaded area in Fig. 1) we have shown that
pressure applied at room temperature facilitates oxygen diffusion and
induces the electronic phase separation. However, if oxygen is frozen at low 
$T$ and pressure is changed below $100\ K$ we expect that the initial (room
temperature) state is preserved. Fig. 9 shows the pressure effect on $\chi
_{ac}(T)$ for $La_{2}CuO_{4.055}$ ($n_{h}=0.09$). In the first experiment
low pressure ($0.1\ GPa$) was applied at room temperature and the sample was
cooled to low $T$ (dotted curve 1 in Fig. 9 A). Then pressure was increased
(below 100 K) to about 0.5 GPa (solid curve 2 in Fig. 9 A). The
susceptibility curve shifts in parallel to higher temperature at a rate of $%
2.9\ K/GPa$. No broadening or splitting of the transition is observed. In
the second experiment we applied $0.8\ GPa$ at room temperature. The
superconducting transition broadens (dotted curve, Fig. 9 B) and indicates
the tendency to splitting into two transitions due to the phase separation
as seen in the previous section (Fig. 5 A and B). The release of pressure at
low $T$ to $0.1\ GPa$ does not change the width or the shape of the
diamagnetic susceptibility signal (curve 2, Fig. 9 B). The phase separation
induced by pressure at high temperature is preserved. However, $T_{c}$\
shifts by about $2$ to $3\ K$. Heating the sample at $0.1\ GPa$ to room
temperature recovers the original (one-transition) state (curve 3, Fig. 9 B
which is comparable with curve 1, Fig. 9 A) due to the rearrangement of the
interstitial oxygen and hole system towards thermal equilibrium.

In Section III B we have shown that in $La_{1.985}Sr_{0.015}CuO_{4.020}$
high pressure (applied at room temperature) induced the $30\ K$
superconducting state at the expense of the $15\ K$ phase. To show that this
effect is also related to the rearrangement of the interstitial oxygen we
measured the pressure effect using the He gas pressure apparatus. The
experiment starts with applying high pressure of $1.1\ GPa$ at ambient
temperature. The ac susceptibility clearly shows the emerging $30\ K$ phase
at low temperature as indicated by the shoulder in the derivative, $d\chi
_{ac}/dT$, close to $30\ K$ (curve 1, dotted line in Fig. 10). Although the
pressure is not high enough to show the larger effect as in Fig. 6 the data
are in good agreement with the clamp cell experiments at the same pressure.
After releasing pressure completely (curve 2 in Fig. 10) the $30\ K$ phase
still survives at about the same amount. Only after heating the sample to
room temperature (curve 3, Fig. 10) the oxygen ions and the holes rearrange
and the $30\ K$ phase disappears again. This effect can only be explained by
considering the pressure induced change of the interstitial oxygen
distribution. High pressure stabilizes the hole-oxygen rich phase with a
higher ($30\ K$) $T_{c}$. After cooling to below $100\ K$ and pressure
release the clustered oxygen ions cannot relax to their zero-pressure
thermal equilibrium distribution. Pinning the holes to the $O^{2-}$-rich
regions the amount of the $30\ K$ phase is preserved. However, this state is
not in global thermal equilibrium. Raising the temperature to ambient the
diffusion of oxygen is initiated and the hole-oxygen system relaxes closer
to equilibrium. Now the hole system is affected by the randomly distributed
and fixed Sr-ions and the $15\ K$ state dominates the superconducting
properties.

\section{SUMMARY AND CONCLUSIONS}

The pressure effect on the superconducting transitions of electrochemically
doped $La_{2-x}Sr_{x}CuO_{4+\delta }$ was investigated for $x=0,\ 0.015$
with various $\delta $. We show a strong correlation between the hole
carriers and the dopants. Cation- and oxygen-doping are characteristically
different. Hard-doping using strontium creates a correlated dopant-hole
state and inhibits the macroscopic phase separation. Soft-doping using
mobile interstitial oxygen, however, facilitates the tendency to phase
separation into hole-rich and hole-poor phases. Accordingly, $%
La_{2-x}Sr_{x}CuO_{4+\delta }$ shows a rich phase diagram as function of
doping. In particular, two characteristic regimes can be distinguished. One
superconducting transition ($T_{c}\thickapprox 30\ K$) is observed in the
low doping regime but two successive transitions at about $15\ K$ and $30\ K$
appear in the higher doping regime. For the two different sets of samples
with $x=0$ and $x=0.015$ the two regimes are separated by a critical hole
density, $n_{h}\thickapprox 0.085$. The Sr co-doped $%
La_{1.985}Sr_{0.015}CuO_{4+\delta }$ shows a correlated superconducting
phase at low hole density ($n_{h}<0.06$) with a lower $T_{c}\thickapprox 15\
K$ due to the hole pinning effect of the Sr ions.

Pressure applied at room temperature increases the $30\ K$ transition
temperature, $T_{c}^{(30)}$, at an unusually high rate of 9 to 10 K/GPa. In
addition, the tendency to electronic phase separation is strongly enhanced
by pressure as indicated by the large positive $dT_{c}^{(30)}/dp$ and a
negative pressure coefficient of the 15 K transition temperature, $%
T_{c}^{(15)}$. It was shown that, for the special hole density $%
n_{h}\thickapprox 0.085$, pressure induces the phase separation into the two
superconducting phases for both strontium contents, $x=0$ and $x=0.015$. The
data are interpreted as a cooperative effect of pressure on the electronic
state facilitating the phase separation and the re-arrangement of the
interstitial oxygen which enhances the original pressure effect. We conclude
that the electronic state of carefully annealed oxygen-intercalated La-214
is closer to thermal equilibrium due to the missing frustration of immobile
dopant ions.

The latter conclusion is further supported by a series of experiments where
pressure was varied at low temperature (below $100\ K$) inhibiting any
oxygen diffusion. The pressure coefficients of $T_{c}$ appeared to be much
smaller ($<3\ K/GPa$) and comparable with the cation-doped $%
La_{2-x}(Sr,Ba)_{x}CuO_{4}$ indicating that the lower $dT_{c}/dp$ is an
effect of charge pinning{\bf .} It was also shown that pressure, if changed
at low temperature, did neither enhance the tendency of phase separation
observed for $n_{h}>0.085$ nor induce phase separation at the special hole
concentration of $n_{h}\thickapprox 0.085$. We suppose that the immobile
interstitial oxygen at low temperature act similarly as the (randomly
distributed) Sr-cations which may explain the reduced pressure effect on the
electronic hole system.

It is interesting to note that the crossover from the single to double
transition region in the phase diagram as well as the pressure effects
discussed above appear at the same hole density for both systems, $%
La_{2}CuO_{4+\delta }$ and $La_{1.985}Sr_{0.015}CuO_{4+\delta }$, considered
in this investigation. This implies that the ''total'' hole density is the
relevant parameter determining the nature of the superconducting states as
well as the pressure effects on it. For higher strontium contents, however,
effects of Coulomb interaction with the Sr ions have to be taken into
account. The results obtained at ambient and high pressure provide strong
evidence for the importance of the interaction of holes and dopants in high
temperature superconductors. Soft- dopants, like interstitial oxygen in
La-214 or chain oxygen in YBCO, may facilitate the phase separation into
hole rich and hole poor phases. Thermodynamically, soft-dopants and holes
should be considered as one global correlated system relaxing towards
thermodynamic equilibrium at high enough temperatures. Hard-dopants,
however, cause pinning of holes and prevent the electronic system from phase
separation into phases with different hole densities.

As a final note, while electronic phase separation in a very low doping
regime can be understood in terms of Emery's picture,\cite{Emery,Kivelson}
the phase separations we observed in such high doping level is especially
peculiar because the system is electronically phase separated into two
dictinct $T_{c}$'s. This suggests that a distinct electronic structure
exists for each different $T_{c}$ and it starts to form at a temperature
above 200 K.


\acknowledgments
This work was supported by the State of Texas through the Texas Center for
Superconductivity of the University of Houston.


\begin{figure}[tbp]
\caption{Phase diagram of La$_{2-x}$Sr$_{x}$CuO$_{4+\protect\delta }.$%
\newline
Typical examples of the real part of $\protect\chi_{ac}$ are shown and the
arrows indicate the hole density in the diagram. The shaded area separates
the two regions where one and two superconducting transitions are observed.}
\label{Fig. 1}
\end{figure}

\begin{figure}[tbp]
\caption{Pressure effect on the real part of the ac susceptibility of $%
La_{2-x}Sr_xCuO_{4+\protect\delta}$ in the one transition region ($n_h<0.085$%
).\newline
(A) $La_2CuO_{4.04}$ and (B) $La_{1.985}Sr_{0.015}CuO_{4.027}$. The dotted
curves show the ambient pressure data.}
\label{Fig. 2}
\end{figure}

\begin{figure}[tbp]
\caption{Pressure dependence of $T_c$ (defined as the onset of the
diamagnetic drop) for the data of Fig. 2 (circles). Also shown in (B) are
data for higher doping, $\protect\delta=0.04$ (triangles) and $\protect\delta%
=0.061$ (squares).\newline
The pressure coefficients given in the figure refer to the linear high
pressure part of $T_c(p)$.}
\label{Fig. 3}
\end{figure}

\begin{figure}[tbp]
\caption{Pressure effect on the real part of the ac susceptibility of $%
La_{2-x}Sr_xCuO_{4+\protect\delta}$ in the two transition region ($n_h>0.085$%
).\newline
(A) $La_2CuO_{4.08}$ and (B) $La_{1.985}Sr_{0.015}CuO_{4.061}$. The dotted
curves show the ambient pressure data.}
\label{Fig. 4}
\end{figure}

\begin{figure}[tbp]
\caption{Comparison of the pressure effect on the superconducting
transitions in $La_{2-x}Sr_xCuO_{4+\protect\delta}$ near the special hole
density $n_h=0.085$. (A) and (B) show that samples with different oxygen
content, $\protect\delta$, but the same hole density behave very similar
under pressure. (A) and (C) compare samples with the same $\protect\delta$
but slightly different $n_h$.}
\label{Fig.5}
\end{figure}

\begin{figure}[tbp]
\caption{Pressure effect on the superconducting transition in $%
La_{1.985}Sr_{0.015}CuO_{4.02}$.\newline
(A) Real part of the ac susceptibility. The 30 K superconducting state grows
under pressure on the expense of the ambient pressure 15 K state.\newline
(B) The derivative $d\protect\chi_{ac}/dT$ revealing the details of the
transition from the $T_c=15$ K to the $T_c=30$ K state.}
\label{Fig. 6}
\end{figure}

\begin{figure}[tbp]
\caption{Low temperature pressure effect on the $T_c=30$ K phase of $%
La_{1.985}Sr_{0.015}CuO_{4.027}$. 1 GPa pressure was applied at room
temperature (curve 1) and released to zero (curve 3) below 100 K. The inset
shows a further decrease of $T_c$ by 2.2 K after the sample was reheated to
room temperature (solid curve).}
\label{Fig. 7}
\end{figure}

\begin{figure}[tbp]
\caption{Low temperature pressure effect on the superconducting state in the
two-$T_c$ region, $La_{1.985}Sr_{0.015}CuO_{4.061}$. 0.8 GPa pressure was
applied at room temperature and pressure was varied between 0.93 GPa and
zero below 100 K. The inset shows the change of the real part of $\protect%
\chi_{ac}(T)$ after the sample was reheated to room temperature (solid
curve).}
\label{Fig. 8}
\end{figure}

\begin{figure}[tbp]
\caption{Low temperature pressure effects on the real part of the ac
susceptibility of $La_2CuO_{4.055}$ ($n_h=0.09$).\newline
(A) Pressure was increased from 0.1 GPa (curve 1) to 0.5 GPa (curve 2) at
low temperature.\newline
(B) Pressure of 0.8 GPa was applied at room temperature (curve 1) and
released to 0.1 GPa at low T (curve 2). The broadening and splitting into
two transitions is preserved after release of pressure. Reheating to room
temperature initiates the oxygen diffusion and the original 0.1 GPa state is
recovered (curve 3, to be compared with curve 1 in (A)).}
\label{Fig. 9}
\end{figure}

\begin{figure}[tbp]
\caption{Low temperature pressure effects on the superconducting phases in $%
La_{1.985}Sr_{0.015}CuO_{4.023}$ ($n_h=0.06$).\newline
Pressure of 1.1 GPa was applied at room temperature (curve 1) and completely
released at low T (curve 2). The pressure induced $T_c=30$ K phase as
indicated by the shoulder near 30 K in the derivative (inset) is preserved
at zero pressure. Reheating to room temperature allows the oxygen to
redistribute and the 30 K phase disappears again (curve 3).}
\label{Fig. 10}
\end{figure}

%
%


\begin{references}
\bibitem{Schoen}  J. H. Sch\"{o}n, Ch. Kloc, and B. Batlogg, Nature 408, 549
(2000).

\bibitem{Hor}  P. H. Hor, H. H. Feng, Z. G. Li, J. F. DiCarlo, S. Bhavaraju,
and A. J. Jacobson, J. Phys. Chem. Solids 57, 1061 (1996).

\bibitem{Vaknin}  D. Vaknin, J. L. Zaretsky, D. C. Johnston, J. E. Schirber,
and Z. Fisk, Phys. Rev. B 49, 9057 (1994).

\bibitem{Statt}  B. W. Statt, P. C. Hammel, Z. Fisk, S.-W. Cheong, F. C.
Chou, D. C. Johnston, and J. E. Schirber, Phys. Rev. B 52, 15575 (1995).

\bibitem{Jorgensen}  J. D. Jorgensen, B. Dabrowski, S. Pei, D. G. Hinks, L.
Soderholm, B. Morosin, J. E. Schirber, E. L. Venturini, and D. S. Ginley,
Phys. Rev. B 38, 11337 (1988).

\bibitem{Ansaldo}  E. J. Ansaldo, J. H. Brewer, T. M. Riseman, J. E.
Schirber, E. L. Venturini, B. Morosin, D. S. Ginley, and B. Sternlieb, Phys.
Rev. B 40, 2555 (1989).

\bibitem{Hundley}  M. F. Hundley, J. D. Thompson, S.-W. Cheong, Z. Fisk, and
J. E. Schirber, Phys. Rev. B 41, 4062 (1990).

\bibitem{Hammel}  P. C. Hammel, A. P. Reyes, Z. Fisk, M. Takigawa, J. D.
Thompson, R. H. Heffner, S.-W. Cheong, and J. E. Schirber, Phys. Rev. B 42,
6781 (1990).

\bibitem{Zolliker}  P. Zolliker, D. E. Cox, J. B. Parise, E. M. McCarron
III, and W. E. Farneth, Phys. Rev. B 42, 6332 (1990).

\bibitem{Radaelli}  P. G. Radaelli, J. D. Jorgensen, A. J. Schultz, B. A.
Hunter, J. L. Wagner, F. C. Chou, and D. C. Johnston, Phys. Rev. B 48, 499
(1993).

\bibitem{Choi}  J. Choi and J. V. Jose, Phys. Rev. Lett. 62, 320 (1989); H.
Kawamura and M. S. Li, Phys. Rev. B 54, 619 (1996)

\bibitem{Blakeslee}  P. Blakeslee, R. J. Birgenau, F. C. Chou, R.
Christianson, M. A. Kastner, Y. S. Lee, and B. O. Wells, Phys. Rev. B 57,
13915 (1998).

\bibitem{Bhavaraju}  S. Bhavaraju, J. F. DiCarlo, I. Yazdi, A. J. Jacobson,
H. H. Feng, Z. G. Li, and P. H. Hor, Mat. Res. Bull. 29, 735 (1994).

\bibitem{Feng}  H. H. Feng, Z. G. Li, P. H. Hor, S. Bhavaraju, J. F.
DiCarlo, and A. J. Jacobson, Phys. Rev. B 51, 16499 (1995).

\bibitem{Emery}  V. J. Emery, S. A. Kivelson, and H. Q. Lin, Phys. Rev.
Letters 64, 475 (1990).

\bibitem{Kivelson}  S. A. Kivelson, V. J. Emery, and H. Q. Lin, Phys. Rev. B
42, 6523 (1990).

\bibitem{Li}  Z. G. Li, H. H. Feng, Z. Y. Yang, A. Hamed, S. T. Ting, P. H.
Hor, S. Bhavaraju, J. F. DiCarlo, and A. J. Jacobson, Phys. Rev. Letters 77,
5413 (1996).

\bibitem{Yomo}  S. Yomo, K. Soga, Z. G. Li, P. H. Hor, and N. Mori, Physica
C 341-348, 1851 (2000).

\bibitem{Fietz}  W. H. Fietz, R. Quenzel, H. A. Ludwig, K. Grube, S. I.
Schlachter, F. W. Hornung, T. Wolf, A. Erb, M. Kl\"{a}ser, and G.
M\"{u}ller-Vogt, Physica C 270, 258 (1996).

\bibitem{Sade1}  S. Sadewasser, Y. Wang, J. S. Schilling, H. Zheng, A. P.
Paulikas, and B. W. Veal, Phys. Rev. B 56, 14168 (1997).

\bibitem{Sade2}  S. Sadewasser, J. S. Schilling, A. P. Paulikas, and B. W.
Veal, Phys. Rev. B 61, 741 (2000).

\bibitem{Sieburger}  R. Sieburger and J. S. Schilling, Physica C 173, 403
(1991).

\bibitem{Klehe}  A.-K. Klehe, C. Looney, J. S. Schilling, H. Takahashi, N.
Mori, Y. Shimikawa, Y. Kubo, T. Manako, S. Doyle, and A. M. Hermann, Physica
C 257 (1996).

\bibitem{Yomo2}  S. Yomo, M. Kawakami, H. H. Feng, Z. G. Li, P. H. Hor, and
N. Mori, Advances in Superconductivity IX, Springer Verlag, Tokyo (1997), p.
81.

\bibitem{Schirber}  J. E. Schirber, W. R. Bayless, F. C. Chou, D. C.
Johnston, P. C. Canfield, and Z. Fisk, Phys. Rev. B 48, 6506 (1993).

\bibitem{Spain}  I. L. Spain and S. Segall, Cryogenics 11, 26 (1971).

\bibitem{Hor2}  P. H. Hor and S. T. Ting, Preprint TCSUH No. 97:128 (1997).

\bibitem{Hor3}  P. H. Hor and Z. G. Li, Physica C 341-348, 1585 (2000).

\bibitem{XXX}  Q. Xiong, ''High Pressure Study on Structural Instabilities
in High Temperature Superconductors'', Ph.D. Dissertation, University of
Houston, 1993.

\bibitem{Liverman}  W. J. Liverman, J. G. Huber, A. R. Moodenbaugh, and Y.
Xu, Phys. Rev. B 45, 4897 (1992).
\end{references}
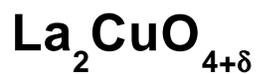
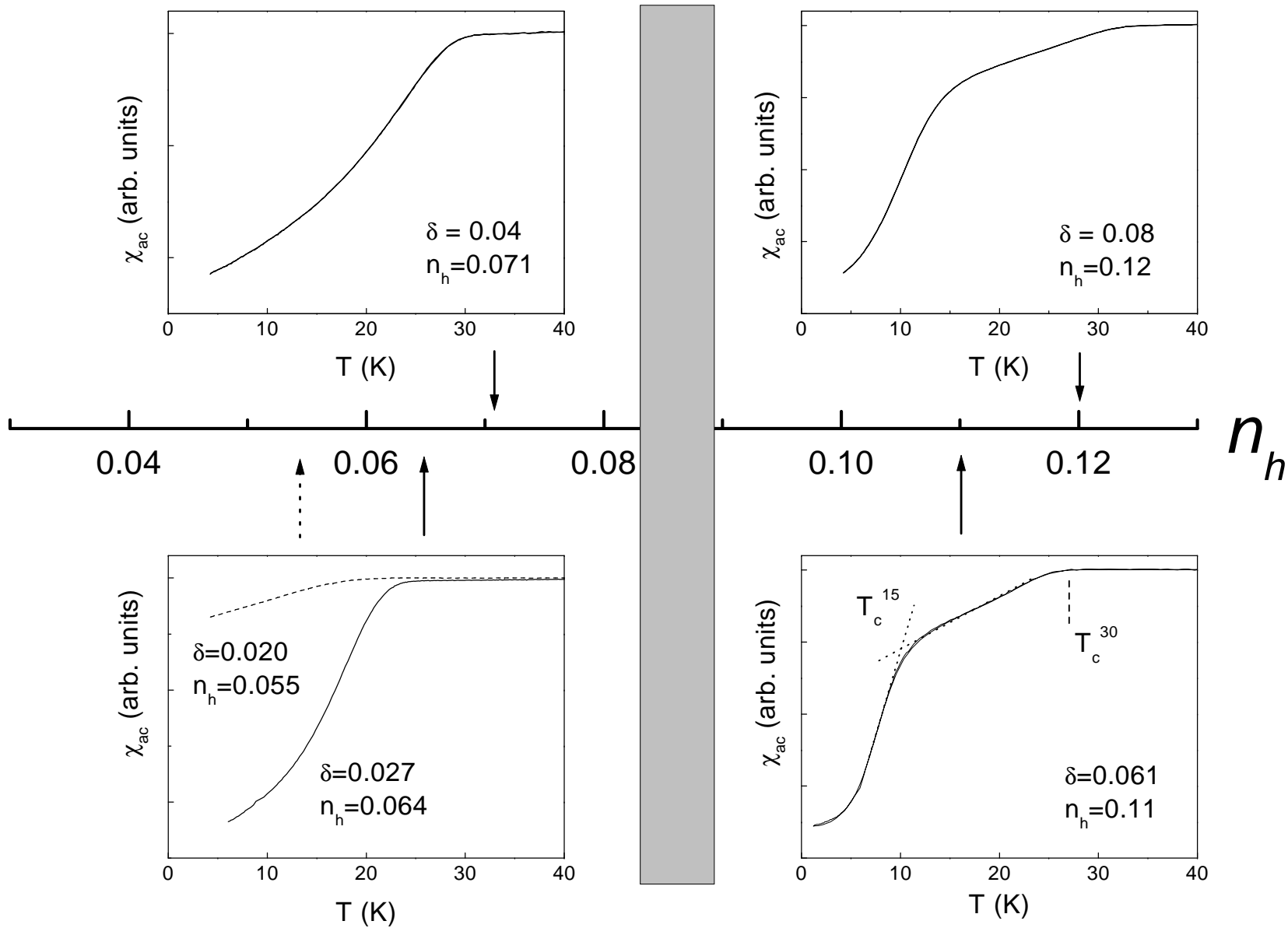
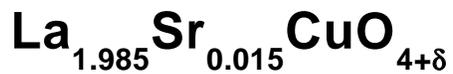
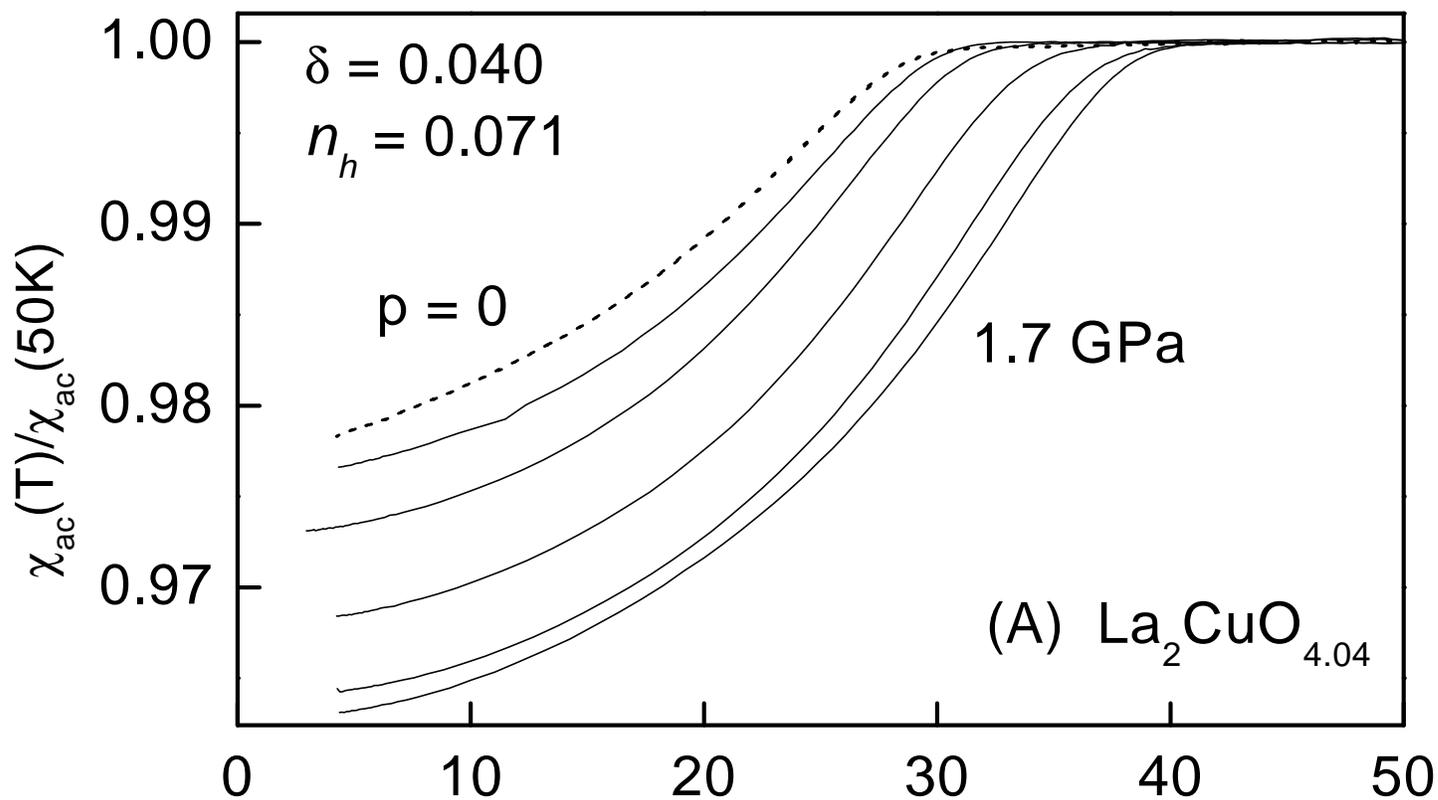
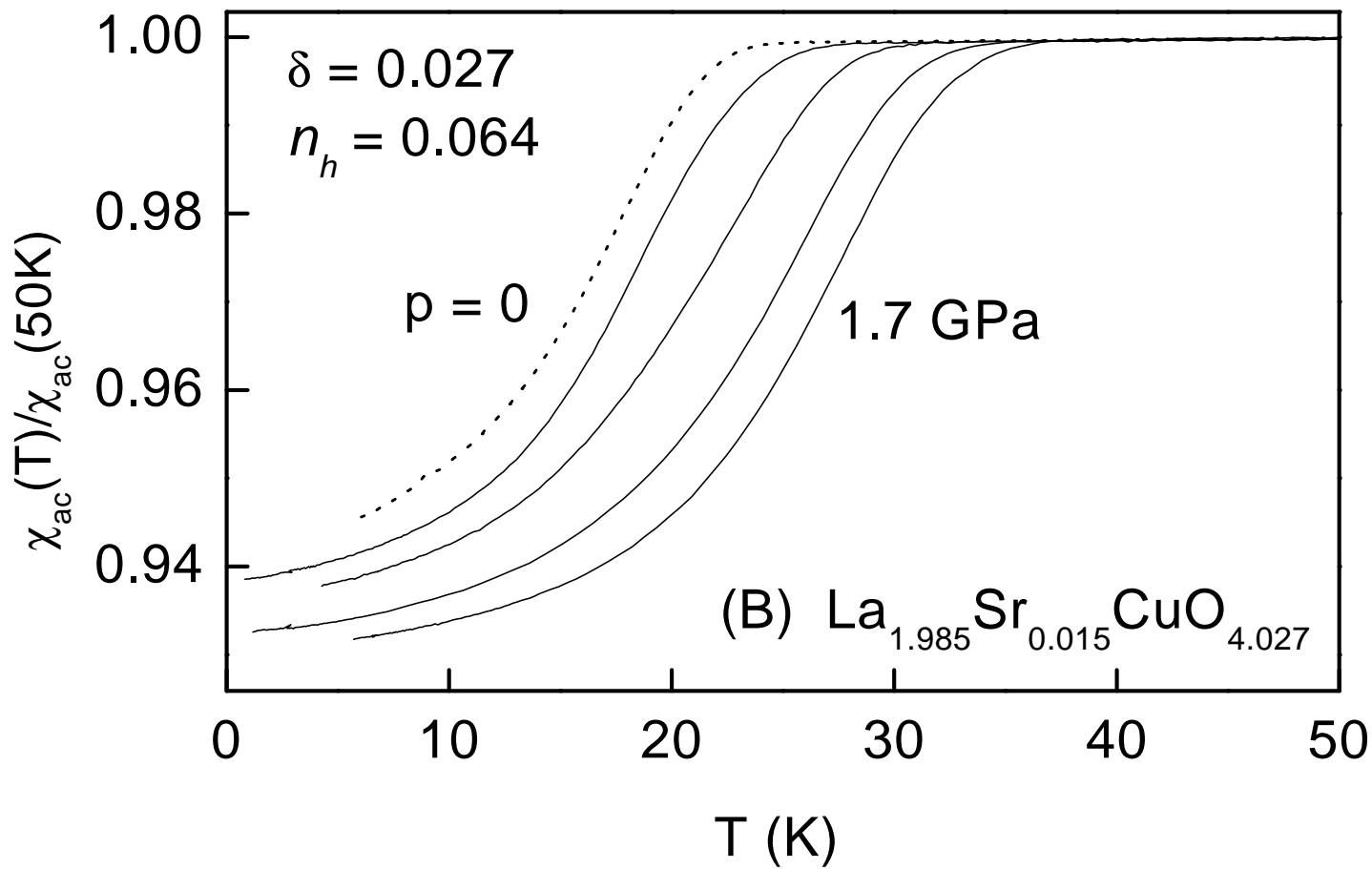
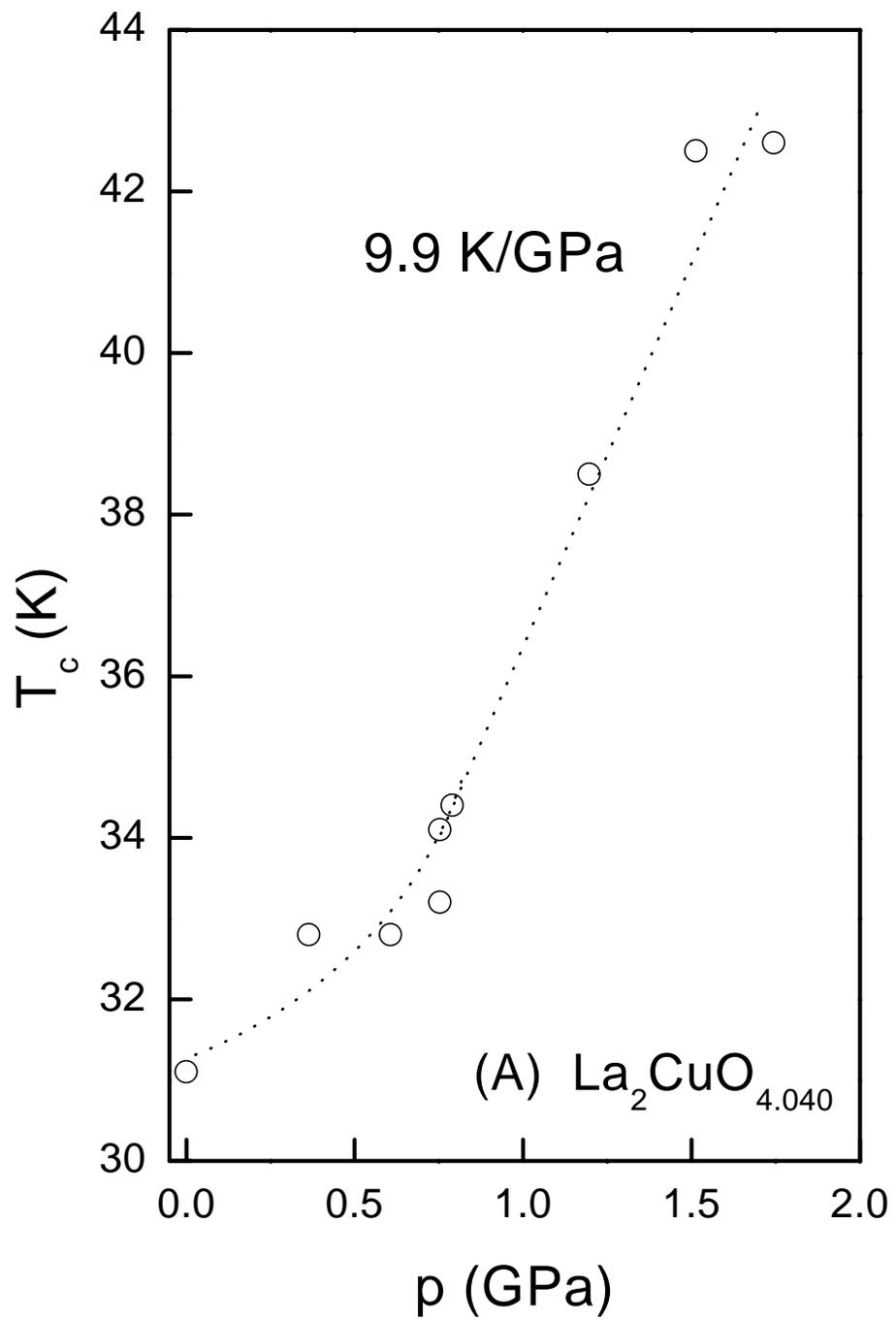
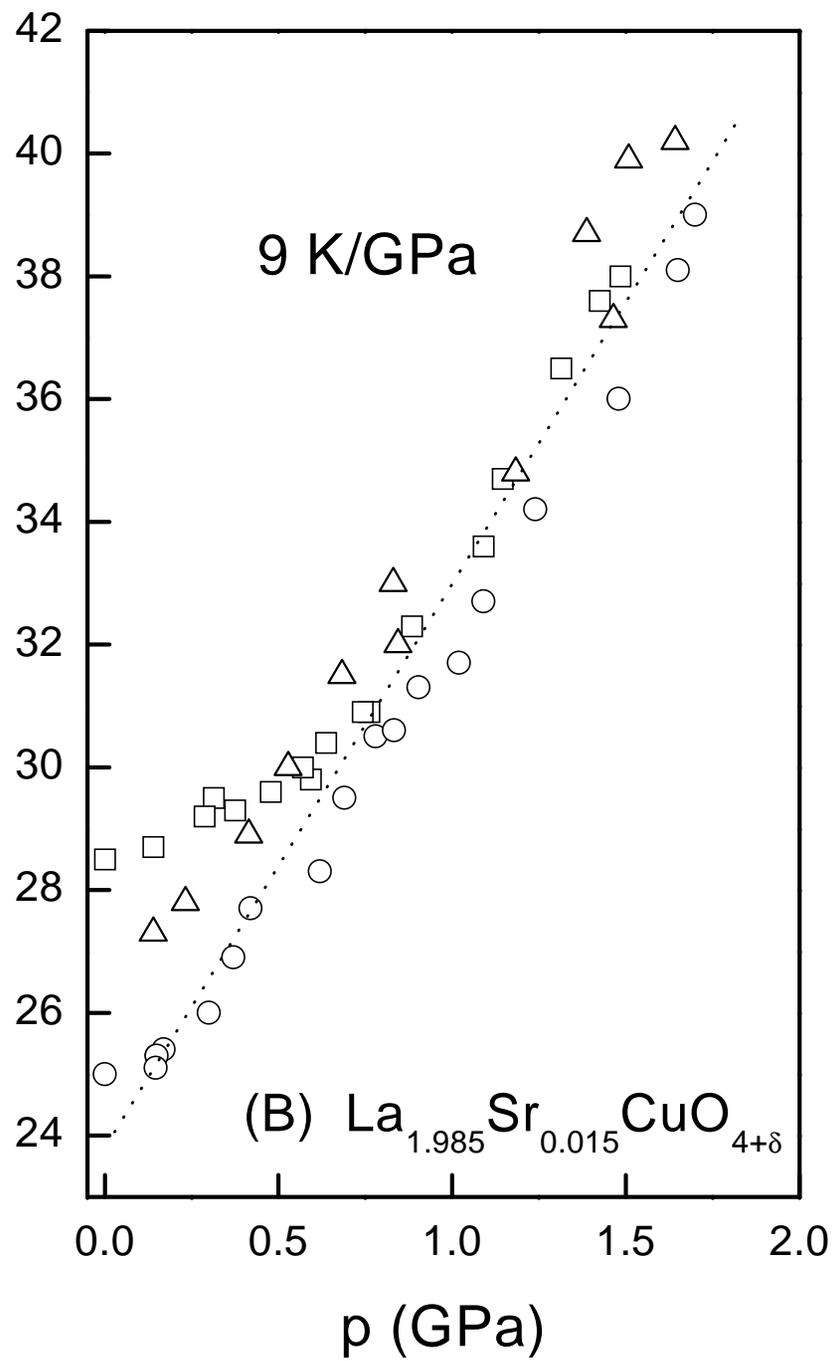
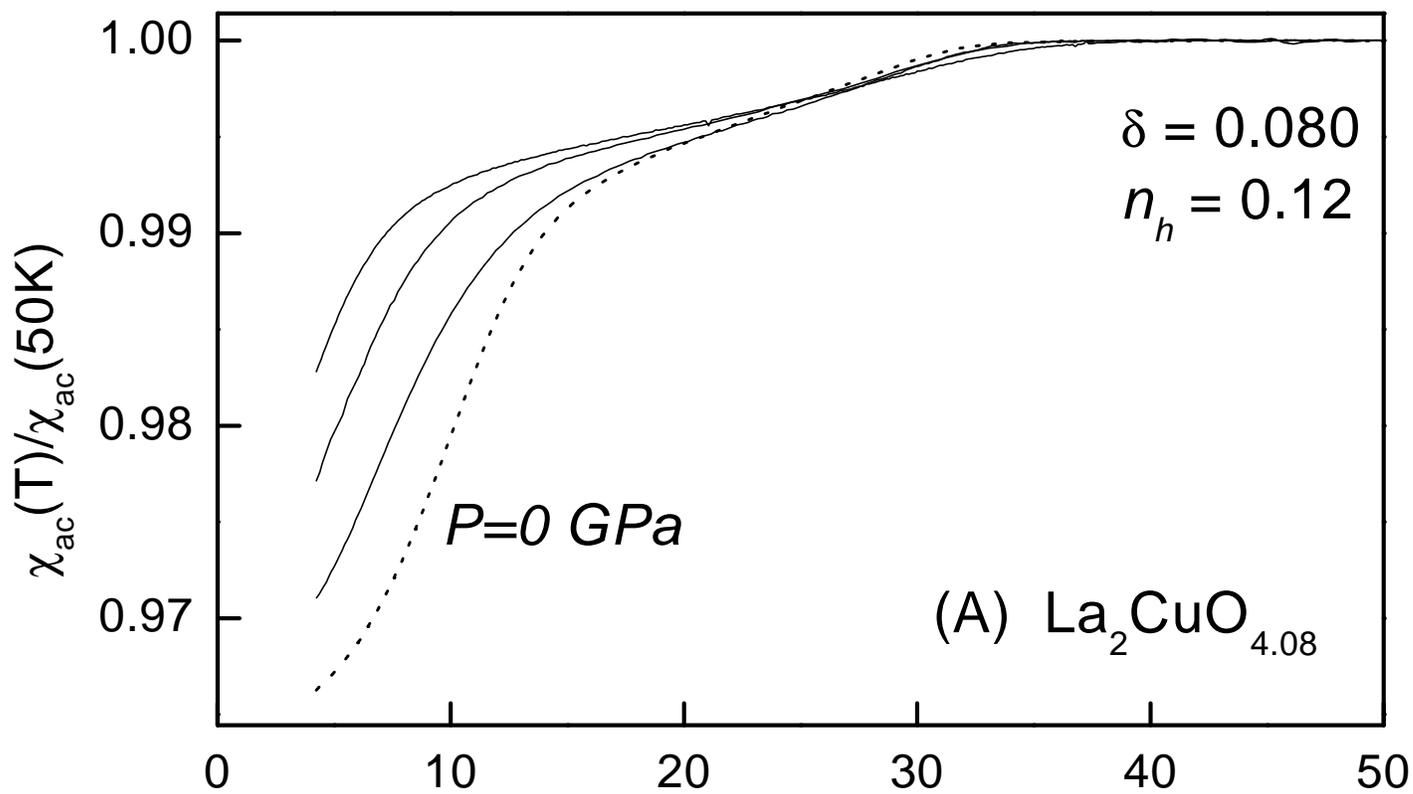
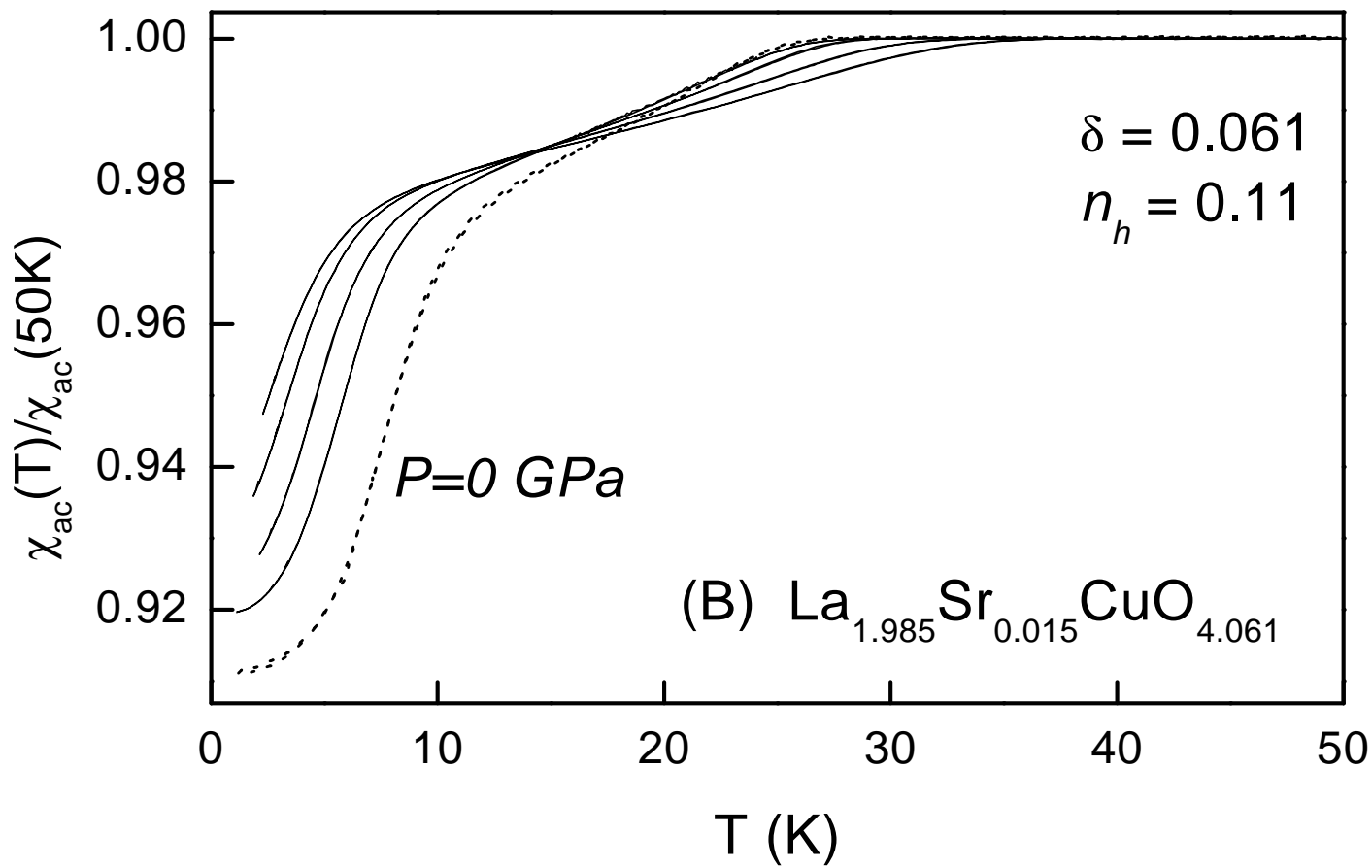
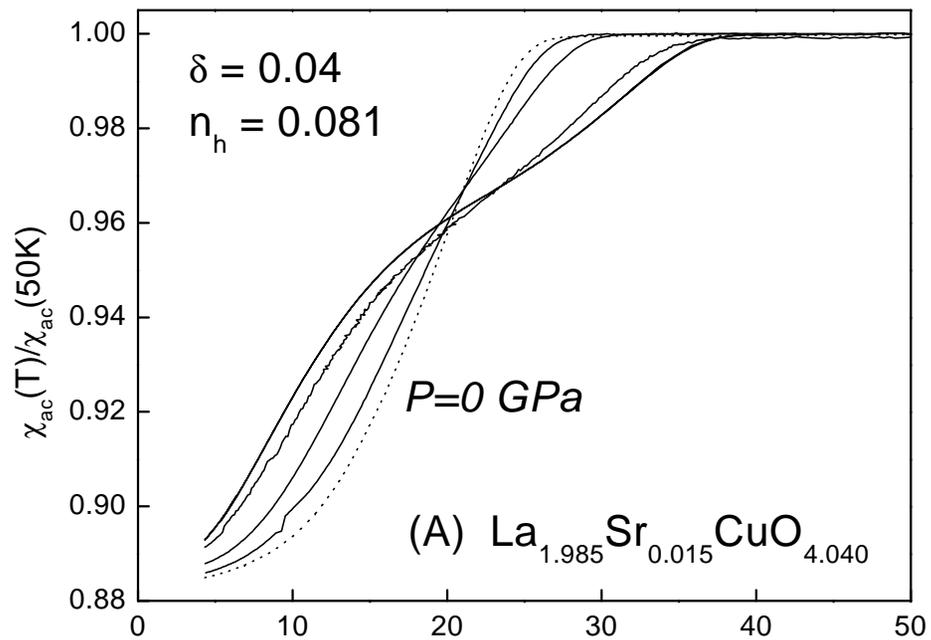
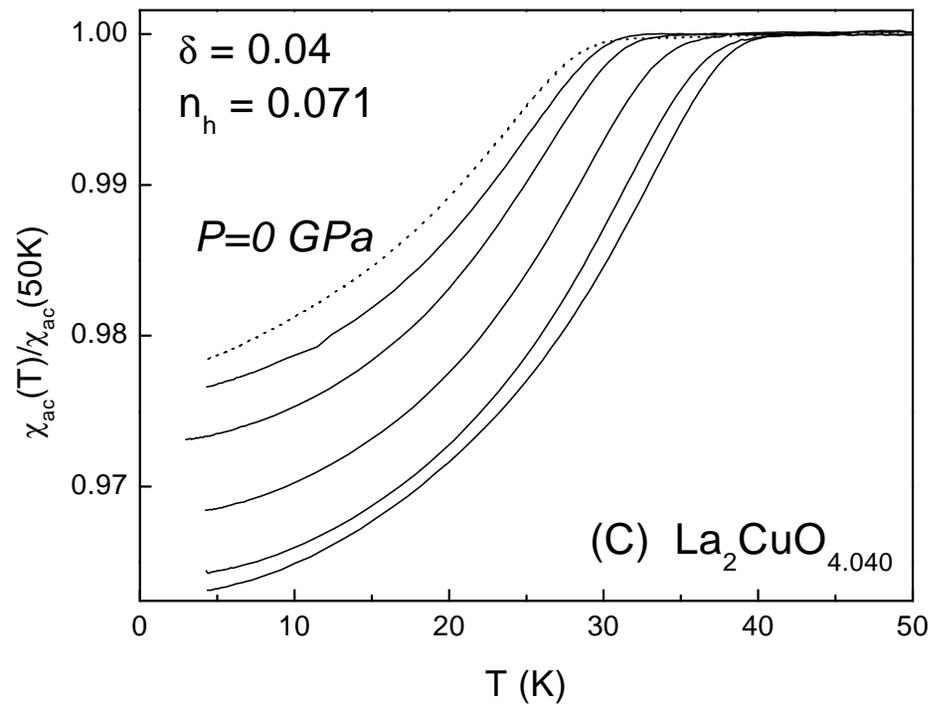
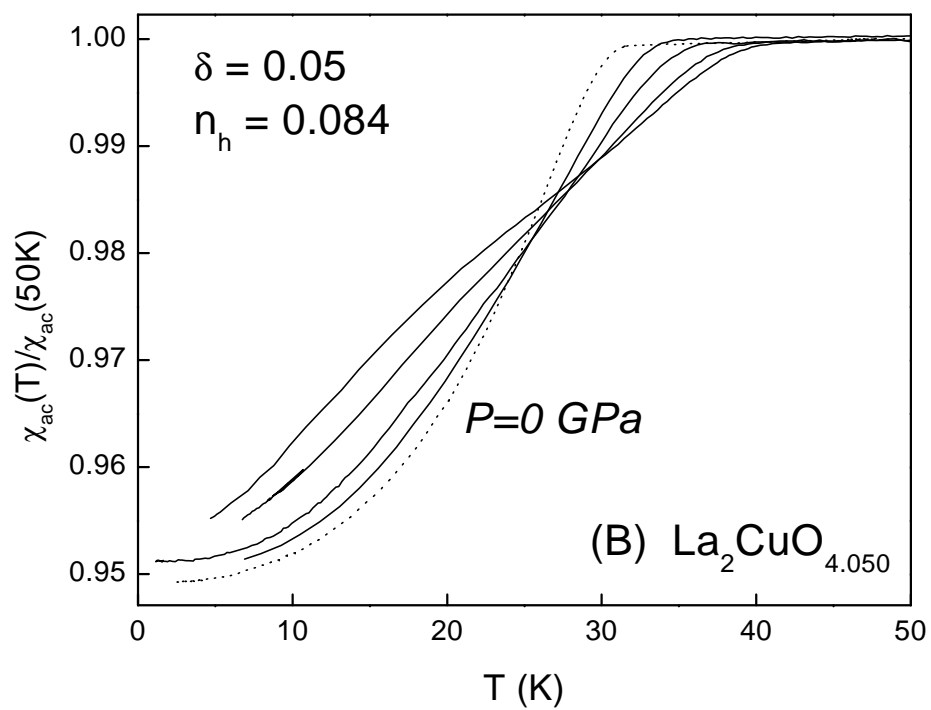
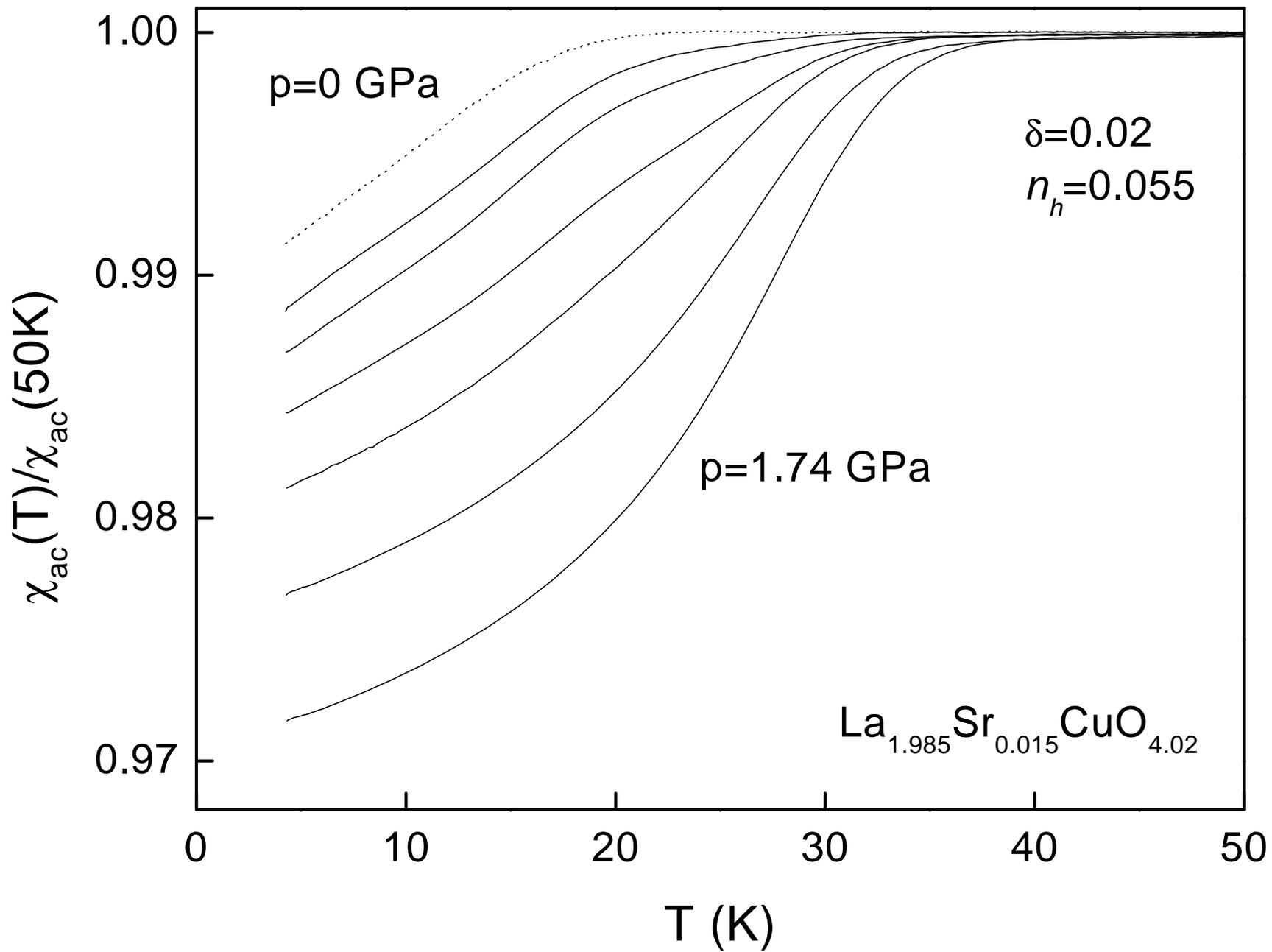
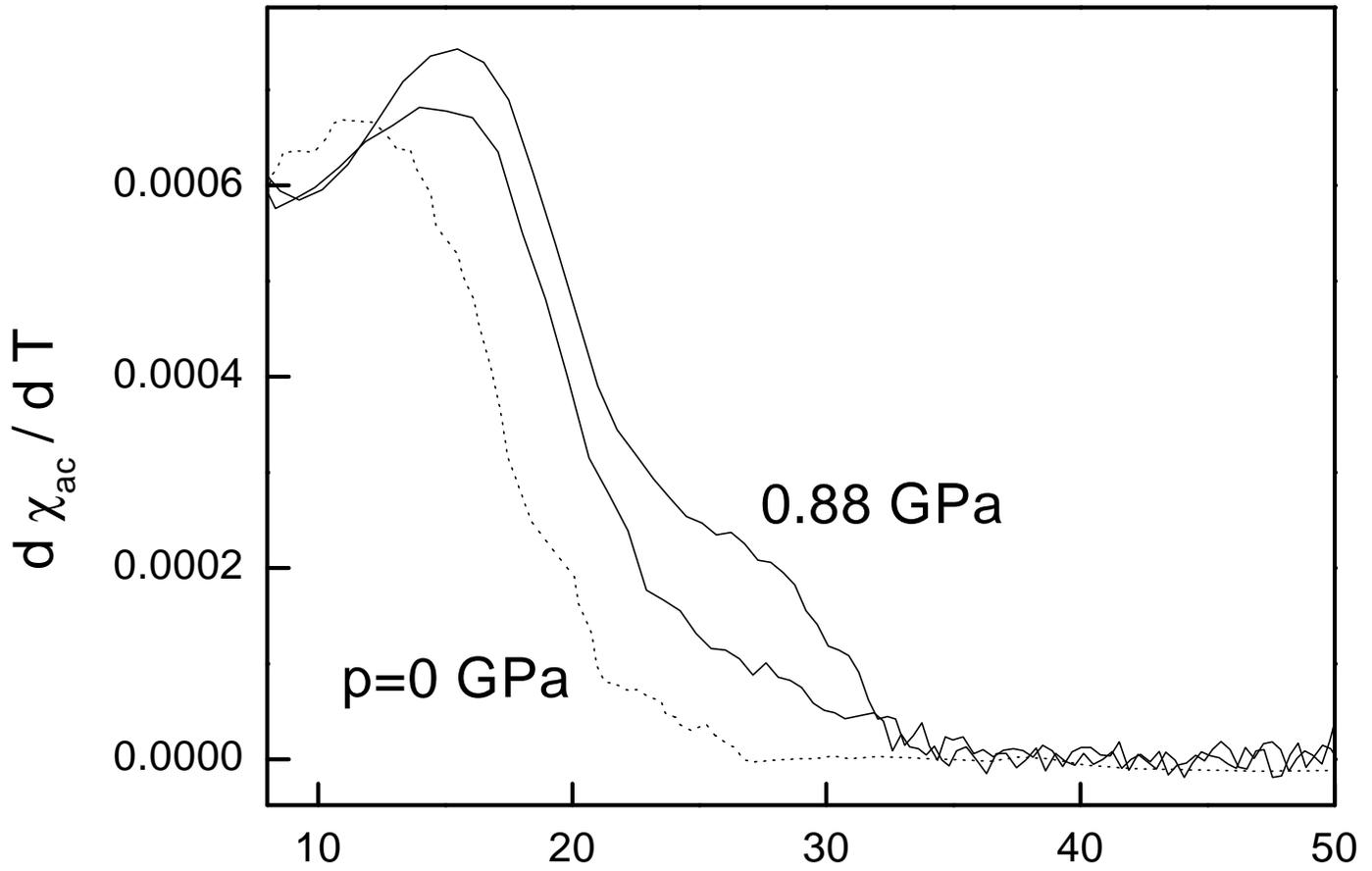
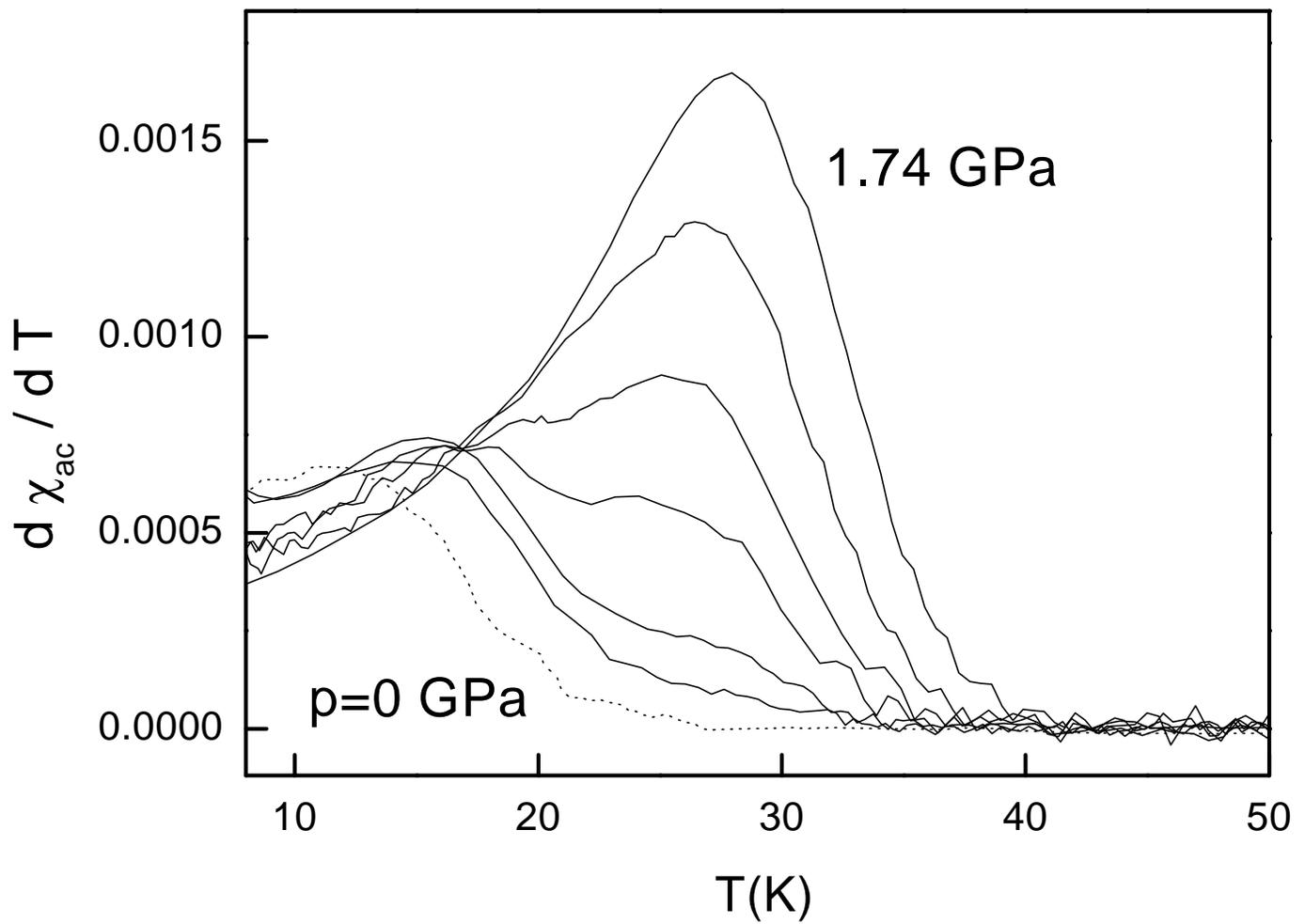
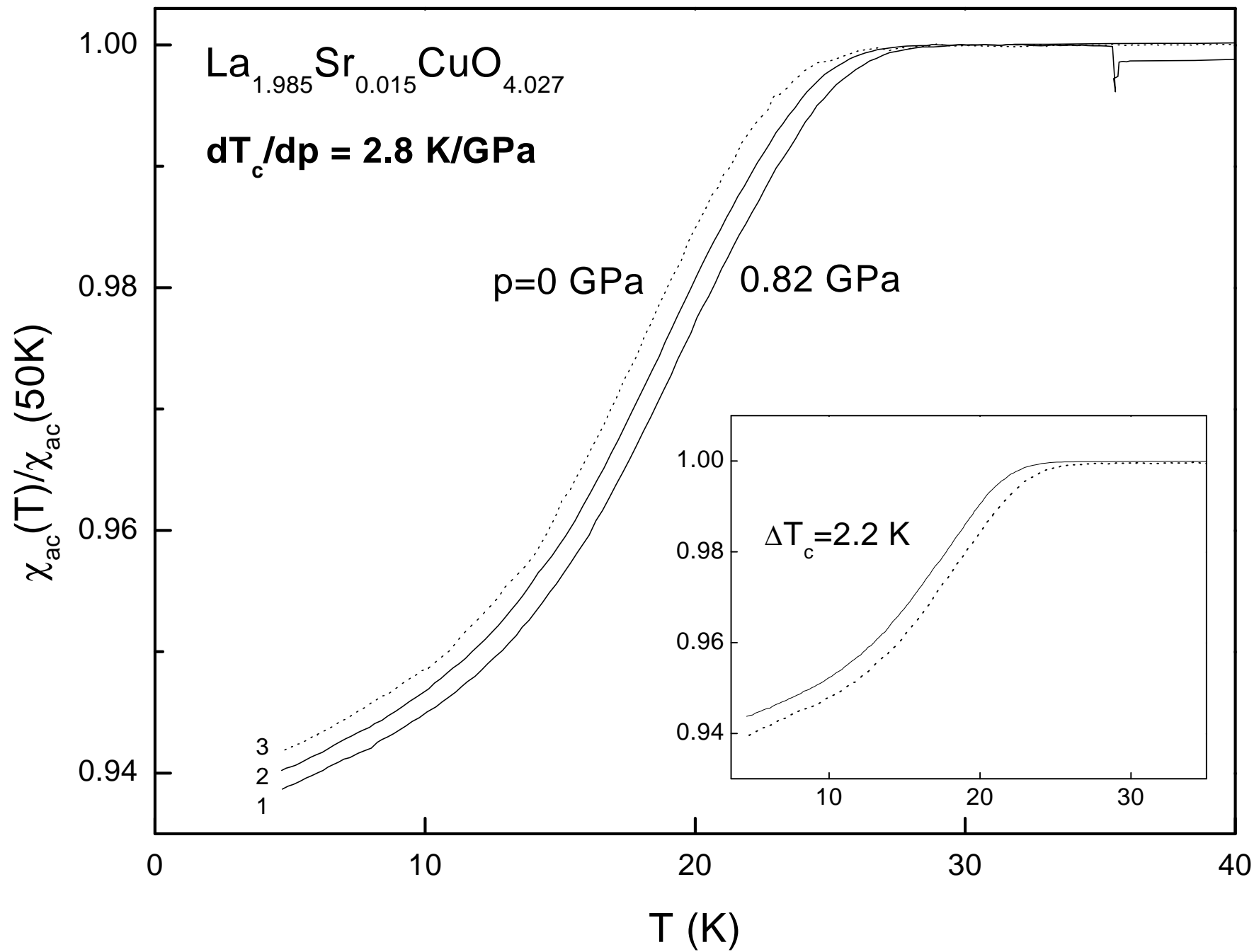
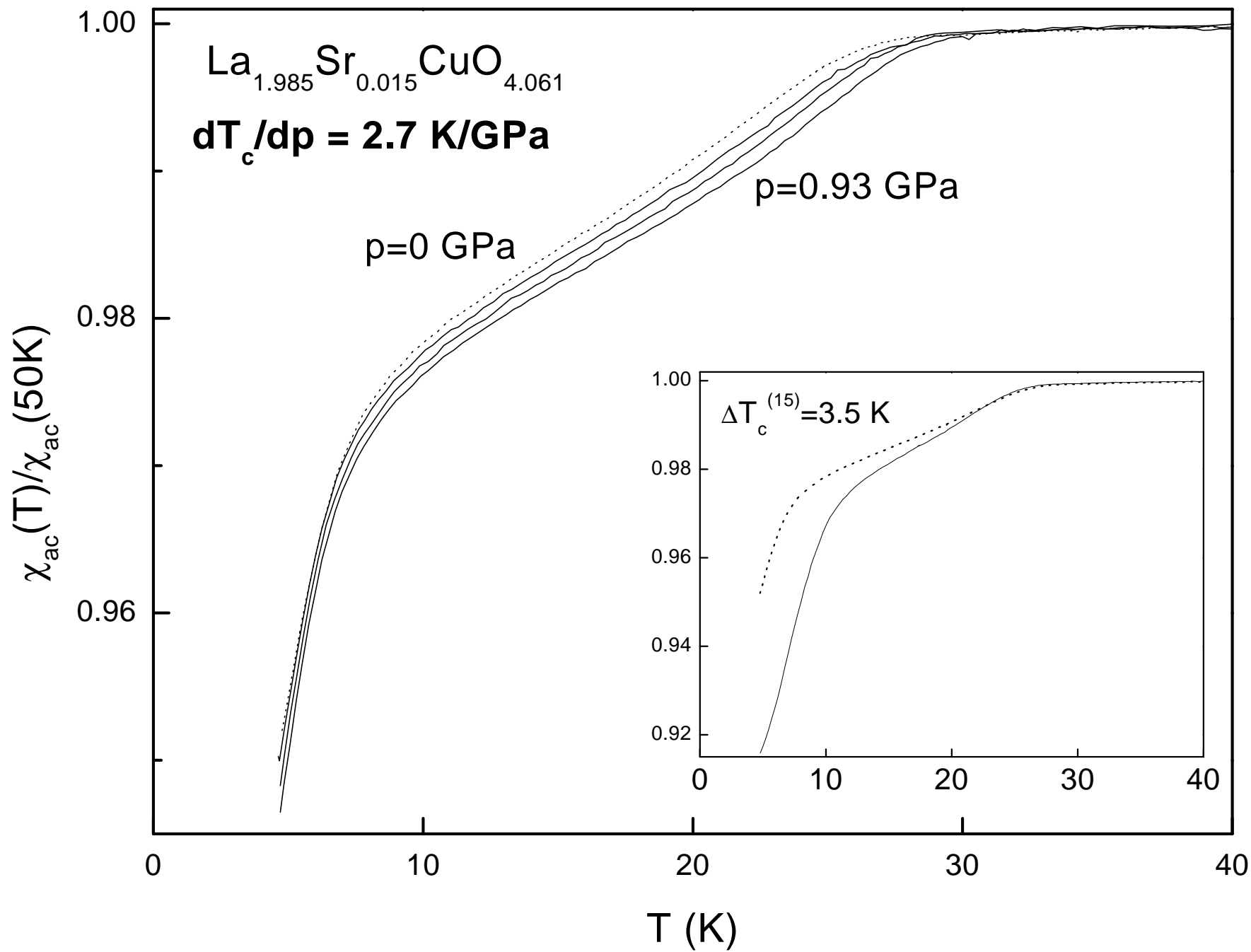
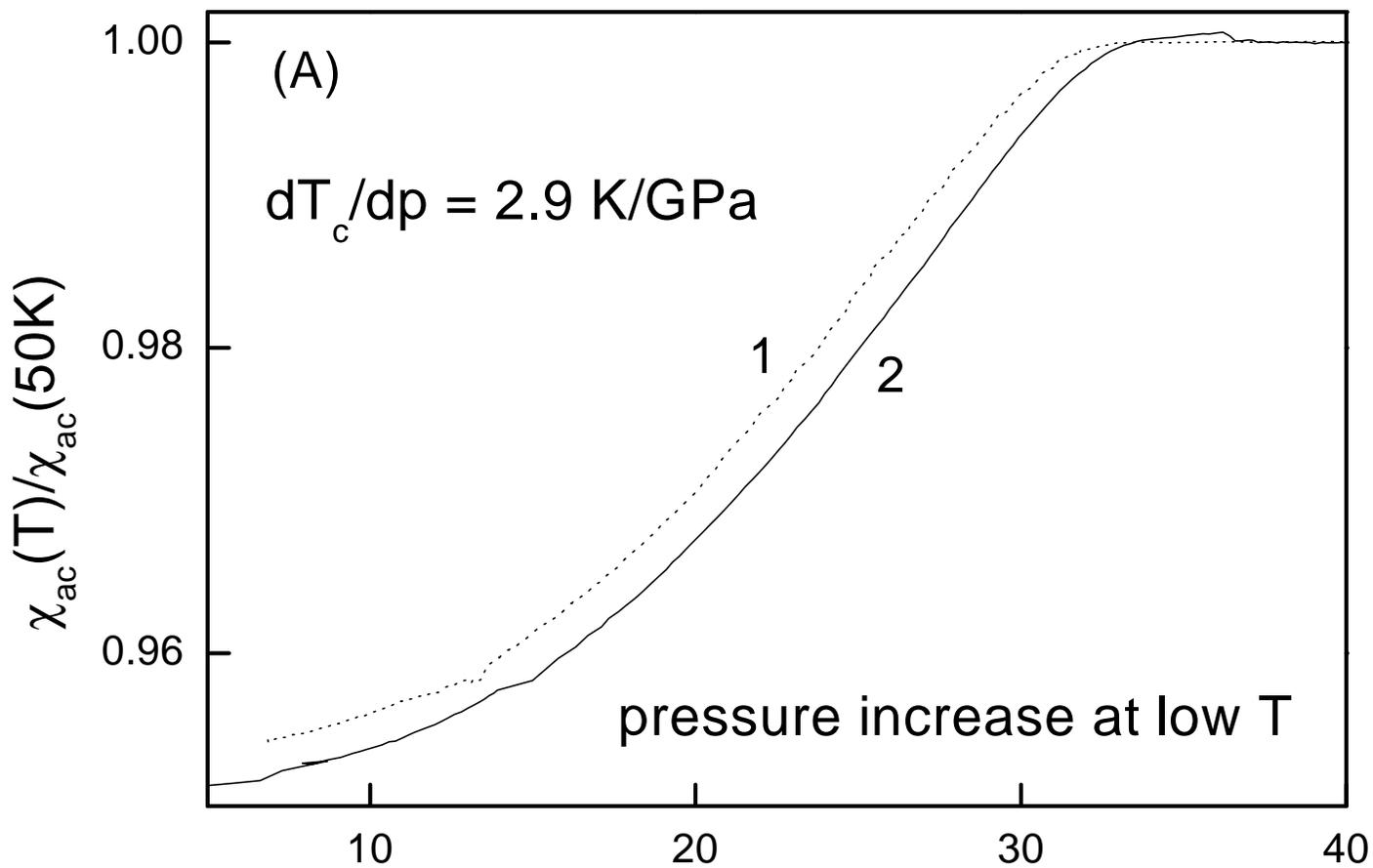
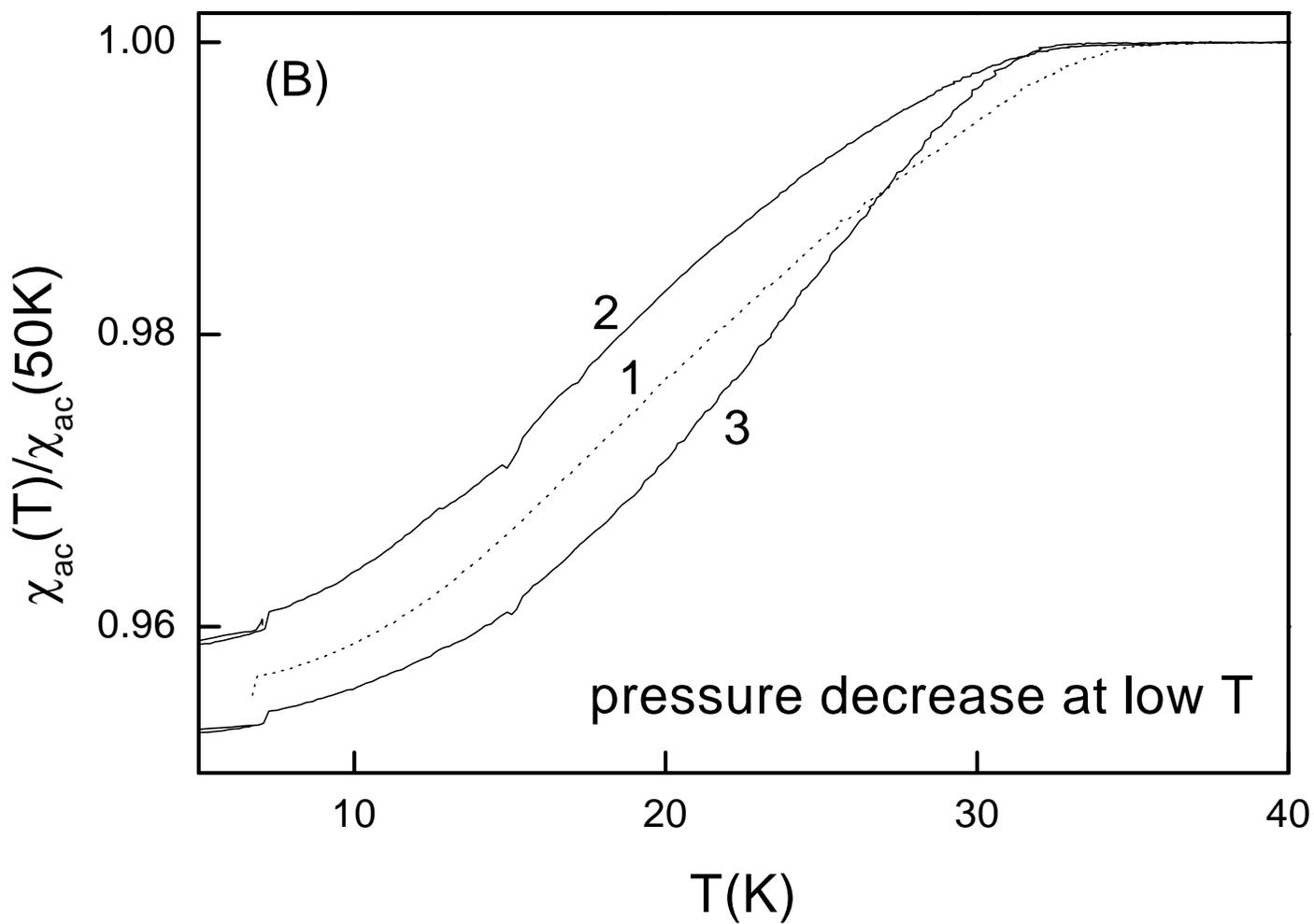
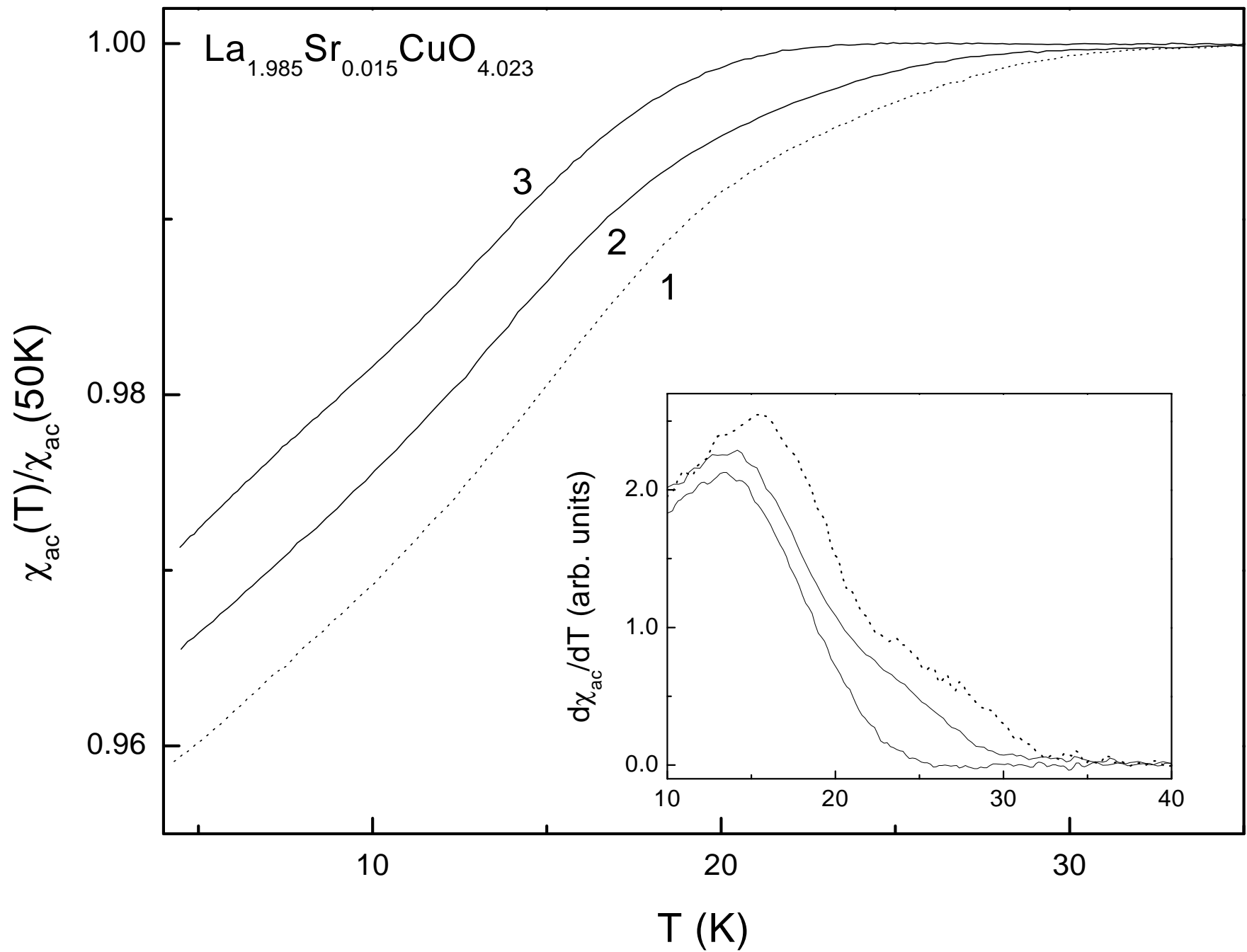
\end{document}